\documentclass[epj,numbook]{svjour}

\usepackage{amsmath,amssymb}
\usepackage{graphicx}
\usepackage{epsf}	
\usepackage{fancyheadings}

\begin{document}
\bibliographystyle{prsty}

\title{A Lattice Gas Coupled to Two Thermal Reservoirs: Monte Carlo and Field
Theoretic Studies}
\author{E. L. Pr\ae stgaard\inst{1} \and B. Schmittmann\inst{2} 
	\and R. K. P. Zia\inst{2,3}}

\institute{Institute for Life Sciences and Chemistry, Roskilde University,
4000 Roskilde, Denmark
\and Center for Stochastic Processes in Science and Engineering,
Department of Physics,\\ 
Virginia Polytechnic Institute and State University, Blacksburg, VA
24061-0435, USA 
\and Fachbereich Physik, Universit\"at - Gesamthochschule Essen,
D-45117 Essen, Germany.}

\date{\today}

\abstract
{\rm We investigate the collective behavior of an Ising lattice gas, driven to
non-equilibrium steady states by being coupled to {\em two} thermal baths.
Monte Carlo methods are applied to a two-dimensional system in which one of
the baths is fixed at infinite temperature. Both generic long range
correlations in the disordered state and critical poperties near the second
order transition are measured. Anisotropic scaling, a key feature near
criticality, is used to extract $T_{c}$ and some critical exponents. On the
theoretical front, a continuum theory, in the spirit of Landau-Ginzburg, is
presented. Being a renormalizable theory, its predictions can be computed by
standard methods of $\epsilon $-expansions and found to be consistent with
simulation data. In particular, the critical behavior of this system belongs
to a universality class which is quite {\em different} from the uniformly
driven Ising model.}


\titlerunning{A Lattice Gas Coupled to Two Thermal Reservoirs: Monte Carlo and Field Theoretic Studies}
\authorrunning {E. Pr\ae stgaard {\em et~al.}}

\maketitle

\section{Introduction}

In recent years, there has been considerable interest in the collective
behavior of many body systems which, despite being far from equilibrium,
have settled in time-independent steady states. Unlike their equilibrium
counterparts, there is no simple Boltzmann factor which provides, in
general, the distribution for these non-equilibrium steady states, so that
most of the progress to date is made by studying specific cases. One
distinguishing feature of these systems is their coupling to more than one
reservoir of energy, so that there is a constant ``{\em through-flux}'' of
energy. Examples include particles driven by external fields such as gravity
or DC electricity, gaining energy from these fields and losing energy
thermally to their environment. An important class is driven diffusive
systems \cite{kls,DL17}, which were motivated physically by the properties
of fast ionic conductors \cite{FIC} and theoretically by their simplicity.
In particular, they are arguably the simplest generalization of the
well-known Ising model \cite{ising} to non-equilibrium conditions. In this
paper, we investigate a lattice gas in which particle hopping rates are
controlled by {\em two thermal baths}, with temperatures $T^{\prime }$ and $%
T $, first introduced in \cite{glms}. Of course, systems subjected to two
(or more) baths are extremely common, from food being cooked on a stove to
the earth as a whole. In most cases, temperature {\em gradients} (in space
or time or both) are present, so that the ``steady'' states are typically
inhomogeneous, controlled by ``external'' parameters with macroscopic length
scales. An excellent example is Rayleigh-B\'{e}nard cells, where macroscopic
length scales are introduced through gravity as well as temperature
gradients. Not surprisingly, large scale properties in such systems can be
quite complex, while universal features may be well hidden. In contrast,
this two-temperature model is macroscopically homogeneous in both space and
time, so that crucial features associated with non-equilibrium steady states
are prominently displayed in the foreground.

Of the many large scale (in both space and time) phenomena displayed by this
class of non-equilibrium systems, two have received considerable attention.
One is the presence of long range correlations at all temperatures above
criticality, despite the absence of long range interactions or dynamics
(e.g., jumps) at the microscopic level. First revealed in the uniformly
driven lattice gas \cite{LRC-DDS}, they have also been observed in the {\em %
randomly} driven lattice gas \cite{LRC-RDDS} and the two-temperature models 
\cite{LRC-TT}. The other remarkable behavior is associated with the critical
point, where significant deviations from the Ising universality class were
found \cite{marro,VM,ktl,epl,wang,antimarro}. Here, we focus on the simple
two-temperature Ising lattice gas and present an extensive Monte Carlo
study, with many results beyond those briefly reported in \cite{epl}. In
this model, particle-hole pairs in the $x$-direction exchange randomly,
simulating coupling to a bath with $T^{\prime }=\infty $. Exchanges in the $%
y $-direction do depend on the energy change in a manner consistent with
coupling to a reservoir at $T$. As we vary $T$, we observe a second order
phase transition, similar to the one found by Onsager \cite{Onsager}, but
located at an approximately 40\% {\em higher} value. A careful study of the
two-point correlations in the disordered phase verifies the long-range
nature ($r^{-2}$) to a much better extent than before \cite{LRC-TT}. These
long-range correlations are ``generic'' \cite{glms,GG} and can be easily
understood \cite{zhls} through a Langevin equation for a continuum field,
which is simply a non-equilibrium version of time-dependent Landau-Ginzburg
theory. Based on this theory, a renormalization group (RG)\ analysis can be
applied to extract the critical behavior. Reported briefly in the past \cite
{SZ-prl,S-epl}, the details of such an analysis are given here. Similar to
the uniformly driven lattice gas, momenta in the $x$- and $y$-directions
scale with different power laws. Unlike the uniformly driven lattice gas,
where the powers differ by a factor of three (in two dimensions), the
difference here is much closer to the ``mean-field'' value of two. Relying
on these results, we carried out extensive simulations using a series of 
{\em rectangular} samples ($L_{x}\sim L_{y}^{2}$), so as to simplify the
finite size scaling analysis. We measured a cumulant ratio, the
``magnetization'', the ``specific heat'' and ``energy'' fluctuations. A
consistent picture emerges, in general agreement with theoretical
predictions. In contrast, using the customary square samples ($L_{x}=L_{y}$%
), no consistent set of parameters can be found to collapse the data.
Undoubtedly, a new scaling variable is playing a dominant role here. Our
conclusion is that anisotropic scaling is crucial and that the simulations
support the field theoretic renormalization group analysis of this system.

The paper is organized as follows. In the next section, the lattice model
and its continuum partner will be described. Simulation methods and results,
as well as comparisons with the theoretical predictions, may be found in
Section 3. For completeness, we present details of the field theoretic
renormalization group analysis (Section 4). A concluding section is devoted
to a summary and an outlook.

\section{The Discrete Model and A Continuum Theory}

\subsection{A Two Temperature Lattice Gas}

The constituents of our model are identical to the two-dimensional Ising
lattice gas \cite{ising,YangLee}. On a square lattice with fully periodic
boundary conditions, the sites $\vec{i}\equiv \left( i_{x},i_{y}\right) $
may be empty or occupied by a single particle, so that a configuration, $%
{\cal C}$, is completely specified by the set of occupation numbers $\left\{
n_{\vec{i}}\right\} $, with each $n_{\vec{i}}$ being $0$ or $1$. In general,
our system will be {\em rectangular}, i.e., 
\begin{eqnarray}
i_{x} & \in & \left[ 1,L_{x}\right] \nonumber \\
i_{y} & \in & \left[ 1,L_{y}\right] \nonumber 
\end{eqnarray}
with $L_{x}\neq L_{y}$ typically. The particles interact with nearest
neighbor attraction, so that, in the Hamiltonian 
\begin{equation}
{\cal H}[{\cal C}]=-4J\sum_{<\vec{i},\vec{j}>}n_{\vec{i}}n_{\vec{j}}
\label{Ham}
\end{equation}
$J$ is positive. We choose $4J$ here so that the effective part of ${\cal H}$
in the spin language is just $-J\sum s_{\vec{i}}s_{\vec{j}}$ , since $%
s=2n-1=\pm 1$. To simulate diffusing particles, we will allow only particle
hops to empty nearest neighbor sites, a dynamics corresponding to Kawasaki
spin exchange \cite{Kawasaki}. Thus, the total particle number is conserved
and, in all cases studies here, fixed to be $N/2$, where 
\begin{equation}
N\equiv L_{x}L_{y}  \label{N}
\end{equation}
Such a half-filled lattice is equivalent to a spin system with zero total
magnetization ($M=0$), so that the critical point can be accessed.

If this system is in contact with a {\em single} thermal bath at temperature 
$T$, then the probability for finding it in any configuration is given by
the Boltzmann factor 
\begin{equation}
P_{eq}[{\cal C}]\propto \exp \left\{ -{\cal H}[{\cal C}]/k_{B}T\right\}
\label{Peq}
\end{equation}
and many of its properties are well known \cite{McCoyWu}. For temperatures
above the Onsager \cite{Onsager} value, $T_{o}\cong 2.2692J/k_{B}$, the
system is in a homogeneous, disordered state. Particle-particle correlations
decay exponentially, governed by a finite correlation length $\xi $. As $%
T\rightarrow T_{o}$, $\xi \rightarrow \infty $ while correlations decay with
a power law. Thermodynamic quantities develop singularities as a function of
various parameters. Due to the constraint $M=0$, the ordered state for $%
T<T_{o}$ is inhomogeneous, consisting of a particle-rich (mainly $s>0$)
region co-existing with a hole-rich one (mainly $s<0$). Given the boundary
conditions, each of these regions is a single strip, of width $L_{>}/2$,
spanning the lattice along $L_{<}$ (where $L_{>}$ is the longer dimension,
etc.). In this manner, the interfacial energy is minimized. Thus, for
example, states with vertical strips will dominate in a system with $%
L_{x}>L_{y}$. Finally, the average $s$ within one of these regions will be,
in the thermodynamic limit, the spontaneous magnetization: $\pm m$.

Our interest in this model lies in its behavior when placed in contact with 
{\em two} baths, at different temperatures. In general, we can expect energy
to flow {\em through} our system, from one bath to another. In the limit
that the baths are infinitely larger than the lattice gas, this energy flux
will settle down to a constant (on the average) and our system reaches a 
{\em non-equilibrium time-independent state}. However, unlike Eqn (\ref{Peq}%
) above, the steady state probability distribution, $P_{ss}[{\cal C}]$ is
not known in general. Indeed, $P_{ss}[{\cal C}]$ is expected to depend on
the details of the dynamics, namely, how our system is coupled to the two
baths. Nevertheless, within a class of models, we believe that there are
many universal properties which do not depend on such details. Here, we will
focus on a particular model, deferring a discussion of the universality
class until Section 4.

To completely specify our model, we list the rules of time evolution:

\begin{itemize}
\item  choose a random nearest neighbor particle-hole pair

\item  if the pair lies along the $x$-direction, exchange them

\item  if the pair lies along the $y$-direction, exchange them with
probability $\min [1,e^{-\Delta {\cal H}/k_{B}T}]$, where $\Delta {\cal H}$
is the energy change due to the exchange.
\end{itemize}

Note that the second rule can be replaced by one similar to the third,
with $T^{\prime }$, the temperature of the second bath. Here, we have
further simplified the model by setting $T^{\prime }=\infty .$ Of course,
this model may be tied ``continuously'' to the equilibrium case by lowering $%
T^{\prime }$ down to $T$.

In our Monte Carlo simulation, a random pair is first chosen. If this is a
particle-hole pair, the rules above apply; otherwise, another pair will be
randomly chosen. A Monte Carlo step (MCS) is defined by $L_{x}L_{y}$
attempts. Starting from, typically, a random initial configuration, the
system is evolved for up to $5\times 10^{6}$ MCS. Allowing the first $10^{5}$
MCS for the system to come to steady state, we measure various quantities
every $10$ - $100$ MCS. The results quoted in the next section are obtained
by averaging these measurements.

As will be discussed in more detail, the second order phase transition is
found to survive. However, the critical temperature is {\em raised} by
approximately $40\%$! In addition, the two-particle correlation function
develops power law decays at all temperatures in the disordered phase, while
the critical properties are modified from the equilibrium Ising class \cite
{WilsonFisher}. To understand such collective behavior in the long-time and
large-scale limit, we often rely on continuum descriptions. This is the
topic of the next subsection.

\subsection{Formulation of a Langevin Equation}

In this subsection, we formulate a continuum approach, based on field
theory. For the convenience of readers who may not be familiar with this
approach, we first summarize the well-established steps for the Ising model
in equilibrium. In that case, a series of continuum theories yield better
and deeper understanding of its properties. The simplest continuum
description starts with Landau's macroscopic free energy function for the
magnetization $M$ 
\[
A(M)=\frac{\tau }{2}M^{2}+\frac{u}{4!}M^{4} 
\]
which focuses only on spatially homogeneous features. To capture spatial
inhomogeneities and thermal fluctuations, we rely on the ``mesoscopic''
Landau-Ginzburg Hamiltonian for a magnetization $\varphi (\vec{x})$ which is
a coarse-grained version of $s_{\vec{i}}$ from the lattice model.
Furthermore, from renormalization group studies, we learned that it is
important to formulate such theories in general dimension $d$ \cite
{WilsonFisher} (even though our simulations are restricted to $d=2$). Thus,
we write 
\begin{equation}
{\cal H}_{LG}{\cal =}\int d^{d}x\{\frac{1}{2}(\vec{\nabla}\varphi )^{2}+%
\frac{\tau }{2}\varphi ^{2}+\frac{u}{4!}\varphi ^{4}\}.  \label{HLG}
\end{equation}
{\em Far from} the critical point, which is modeled by vanishing $\tau $
here, good agreement with data can be gleaned from ${\cal H}_{LG}$ by
inserting Eqn (\ref{HLG}) into Eqn (\ref{Peq}), i.e., $P_{eq}[\varphi (\vec{x%
})]\propto \exp \left\{ -{\cal H}_{LG}\right\} $, and relying on simple
approximation schemes in subsequent computations. Near the phase transition,
however, more powerful renormalization group techniques \cite{RG} are
needed, providing excellent predictions of long-wavelength properties, such
as critical behavior. Further, these approaches can be generalized to
dynamic phenomena \cite{ZJ,DF} near equilibrium, starting with a Langevin
equation for $\varphi (\vec{x},t)$.

For later reference, we briefly review this procedure for the Ising model,
in the presence of a conservation law on the total magnetization $\int
d^{d}x\,\varphi (\vec{x},t)$. Then, the Langevin equation takes the form of
a continuity equation: 
\begin{equation}
\partial _{t}\varphi (\vec{x},t)=-\vec{\nabla}\cdot \vec{J}(\vec{x},t)
\label{CE}
\end{equation}
The form of the current $\vec{J}$ is postulated, guided by the symmetries
and key physical features of the microscopic rates. It typically consists of
a deterministic term, capturing the thermodynamic forces acting on $\varphi $%
, and a Gaussian noise term which reflects the effect of thermal
fluctuations. For the Ising model, the deterministic part ensures that $%
\varphi $ relaxes towards configurations of lower energy. Moreover, $\vec{J}$
must be chosen such that the {\em known} equilibrium distribution, $%
P_{eq}[\varphi (\vec{x})]$, controls the static properties of $\varphi (\vec{%
x},t)$. This constraint forces the Langevin equation to satisfy the
fluctuation-dissipation theorem (FDT) \cite{FDT}. Let us briefly recall the
resulting structure. The most general form of the Langevin equation is 
\begin{equation}
\partial _{t}\varphi ={\Bbb F}(\varphi ,\vec{\nabla}\varphi ...)+\zeta 
\end{equation}
Here, ${\Bbb F}$ is a functional of $\varphi $ and its derivatives,
constructed on physical grounds in the spirit of a Landau expansion. It
reflects the deterministic part of the dynamics. $\zeta $ denotes the noise
term, with correlations 
\begin{equation}
\left\langle \zeta (\vec{x},t)\zeta (\vec{x}^{\prime },t^{\prime
})\right\rangle =2{\Bbb N}\Theta \delta (\vec{x}-\vec{x}^{\prime })\delta
(t-t^{\prime })  \label{GN}
\end{equation}
$\Theta $ is just a positive constant which measures the strength of the
correlations. More importantly, the noise ``matrix'' ${\Bbb N}$ carries
information about conservation laws or internal symmetries which are obeyed
by the dynamics. For our purposes, it is sufficient to consider only a
scalar order parameter $\varphi $ and noise matrices which do not depend on $%
\varphi $. (More general cases are discussed in, e.g., \cite{risken} and 
\cite{hkj}.) Given this basic form, a steady-state solution for the
configurational probability is easily found \cite{risken} {\em provided} $%
{\Bbb F}$ is ``Hamiltonian'', i.e., if it can be written as ${\Bbb N}$
acting on the functional derivative of an appropriate Hamiltonian ${\cal H}$%
: 
\begin{equation}
{\Bbb N}\frac{\delta {\cal H}}{\delta \varphi }\text{ .}  \label{FDT}
\end{equation}
In this case, the steady state is simply the equilibrium distribution, $\exp
(-{\cal H}/\Theta {\cal )}$, irrespective of the choice of ${\Bbb N}$. We
will refer to such a dynamics, in an operational sense \cite{DL17}, as
``FDT-satisfying''. Note that the ``noise strength,'' $\Theta $, plays the
role of a temperature here. Since it just sets the scale for the energy, it
may be absorbed into the definition of ${\cal H}$. In the Ising case, the
microscopic dynamics is isotropic (at least near the critical point) and
order-parameter conserving, so that ${\Bbb N}\propto -\nabla ^{2}$. Our
preceding discussion suggests the equation of motion 
\begin{equation}
\partial _{t}\varphi (\vec{x},t)=\lambda \nabla ^{2}\frac{\delta {\cal H}%
_{LG}}{\delta \varphi }+\zeta (\vec{x},t)  \label{MBa}
\end{equation}
with 
\begin{eqnarray}
\left\langle \zeta (\vec{x},t)\right\rangle &=&\,0  \nonumber \\
\left\langle \zeta (\vec{x},t)\zeta (\vec{x}^{\prime },t^{\prime
})\right\rangle &=&-2\lambda \nabla ^{2}\delta \left( \vec{x}-\vec{x}%
^{\prime }\right) \delta \left( t-t^{\prime }\right)   \label{MBb}
\end{eqnarray}
The current is therefore $\vec{J}(\vec{x},t)=-\lambda \vec{\nabla}\left(
\delta {\cal H}_{LG}/\delta \varphi \right) +\vec{\eta}(\vec{x},t)$, where
the second term is the noisy part, related to $\zeta $ simply by $\zeta (%
\vec{x},t)=-\vec{\nabla}\cdot \vec{\eta}(\vec{x},t)$. The coefficient $%
\lambda $ sets the time scale. In fact, symmetry considerations alone would
have {\em forced} us into this form, even without invoking the FDT: the
deterministic part of (\ref{MBa}), $\nabla ^{2}\left( \delta {\cal H}%
_{LG}/\delta \varphi \right) $ just contains the leading terms of a general
Landau expansion, in powers of $\varphi $ and $\vec{\nabla}\varphi $ which
are consistent with all symmetries of the Ising model.

This Langevin equation, known as Model B \cite{HH}, describes the critical
dynamics of a conserved scalar order parameter. The time-{\em independent},
or static, aspects of its critical behavior are controlled by ${\cal H}_{LG}$
and fall into the equilibrium Ising class. In addition, one obtains
time {\em dependent} properties as well, 
such as the dynamic critical exponent $z$
and the scaling behavior of {\em dynamic} correlation and response
functions. These are characteristic for the Model B class. We note, for
later reference, that the critical parameter, $\tau$, now plays the role of
a diffusion coefficient: As $T$ drops below $T_{c}$, $\tau$ becomes negative
indicating that ``antidiffusion'', i.e., phase segregation, occurs.

Following these lines, we formulate a continuum theory for our {\em %
non-equilibrium} Ising lattice gas. Once again, we begin with a continuity
equation for $\varphi (\vec{x},t)$, so that the conservation law is
guaranteed. Before writing an expression for the current, however, it is
paramount that we should summarize the symmetries of our microscopic theory.
Like the equilibrium Ising model, it exhibits full translation and
reflection invariance, as well as the characteristic Ising ``up-down''
symmetry. However, in contrast to its equilibrium counterpart, even near
criticality we can no longer expect full rotational symmetry, since distinct
temperatures, $T_{\alpha }$, control exchanges along different directions $%
\alpha =1,2,...,d$. In the simulations, $d=2$ and $T_{1}\left( =T_{x}\right)
=T^{\prime }=\infty $ while $T_{2}\left( =T_{y}\right) =T<\infty $. Clearly,
we enjoy some freedom here in extending this model to general dimensions: in
principle, we could imagine couplings to $d$ different temperature baths,
with temperatures $T_{1}\geq T_{2}\geq ...\geq T_{d}$. However, the most
interesting and fundamental case remains that of just two baths, with $%
T_{2}=T_{3}=...=T_{d}=T$ and $T_{1}=T^{\prime }>T$. In other words, we
select a {\em one-dimensional} subspace to be coupled to the higher
temperature ($T^{\prime }=\infty $ in simulations). The full rotational
symmetry, ${\cal O}(d)$, of the Ising model is obviously reduced to an $%
{\cal O}(d-1)$ symmetry in the remaining $(d-1)$ dimensions. We will refer
to the subspace associated with the higher temperature as ``parallel''
(e.g., $T_{\Vert }=T^{\prime }$) and the complementary subspace as
``transverse'' ($T_{\bot }=T$). In the section on field theoretic studies,
we will return briefly to the general case, in order to show that all {\em %
nontrivial} features are indeed captured by the simpler two-temperature
theory.

Given the restricted rotational invariance of our model, we can no longer
expect the {\em same} Landau expansion for the currents in the parallel and
the transverse subspaces. In particular, all $\vec{\nabla}$ operators should
be split into parallel ($\partial $) and transverse ($\vec{\nabla}_{\bot }$)
parts, accompanied by different coefficients. Note, however, that invariance
under reflections guarantees that gradients still appear in pairs: $\nabla
_{\bot }^{2}$ or $\partial ^{2}$. Thus, $\tau \nabla ^{2}$ ``splits'' into $%
\tau _{\bot }\nabla _{\bot }^{2}+\tau _{\Vert }\partial ^{2}$, $u\nabla ^{2}$
becomes $u_{\bot }\nabla _{\bot }^{2}+u_{\Vert }\partial ^{2}$, and $%
\nabla ^{4}$ turns into $\alpha _{\bot }\nabla _{\bot }^{4}+2\alpha _{\times
}\partial ^{2}\nabla _{\bot }^{2}+\alpha _{\Vert }\partial ^{4}$. By
appropriate rescaling of both parallel and transverse lengths, $\alpha
_{\bot }$ and $\alpha _{\Vert }$ can be set to unity. Thus, we can write the
most general Landau expansion which still respects all symmetries of the
microscopic theory as 
\begin{eqnarray}
\partial _{t}\varphi (\vec{x},t) &=& -\vec{\nabla}\cdot \vec{J}(\vec{x},t) 
\nonumber \\
&=&\lambda \{(\tau _{\bot }-\nabla _{\bot }^{2})\nabla _{\bot }^{2}\varphi
+(\tau _{\Vert }-\partial ^{2})\partial ^{2}\varphi \nonumber \\
&\,& -2\alpha _{\times}\partial ^{2}\nabla _{\bot }^{2}\varphi
+ \frac{u}{3!}(\nabla _{\bot }^{2}+\alpha _{3}\partial ^{2})\varphi ^{3}\}
\nonumber \\
&\,& -\vec{\nabla}\cdot \vec{\eta}{\bf (}\vec{x},t)  \label{TTa}
\end{eqnarray}
where, for later convenience, we have written ($u,u\alpha _{3}$) for ($%
u_{\bot },u_{\Vert }$). Turning to the noise, while keeping it Gaussian, we
should expect different variances in the two subspaces. Hence, 
\begin{equation}
\left\langle \eta _{\alpha }(\vec{x},t)\eta _{\beta }(\vec{x}^{\prime
},t^{\prime })\right\rangle =2\lambda \sigma _{\alpha }\delta _{\alpha \beta
}\delta (\vec{x}-\vec{x}^{\prime })\delta (t-t^{\prime })  \label{TTb}
\end{equation}
where the $\sigma $'s represent the noise strengths. Since the transverse
subspace is still fully isotropic, we have $\sigma _{2}=...=\sigma
_{d}\equiv \sigma _{\bot }$. Generically, however, we should expect $\sigma
_{1}\equiv \sigma _{\Vert }\neq $ $\sigma _{\bot }$. Clearly, the noise
terms could also have been expressed via $\zeta (\vec{x},t)\equiv -\vec{%
\nabla}\cdot \vec{\eta}{\bf (}\vec{x},t)$, with correlations $\left\langle
\zeta (\vec{x},t)\zeta (\vec{x}^{\prime },t^{\prime })\right\rangle
=-2\lambda \left( \sigma _{\Vert }\partial ^{2}+\sigma _{\bot }\nabla _{\bot
}^{2}\right) \delta (\vec{x}-\vec{x}^{\prime })\delta (t-t^{\prime })$.
Higher derivatives or powers of the order parameter which have been
neglected above are expected to be small above the critical point. Near $%
T_{c}$, they will be irrelevant, in the renormalization group sense, as we
will see in Section 4.2.1.

This Langevin equation forms the basis of our theoretical analysis. Since
the isotropic diffusion term in Eqn (\ref{MBa}), $\tau \nabla ^{2}\varphi $,
has been modified to $(\tau _{\Vert }\partial ^{2}+$ $\tau _{\bot }\nabla
_{\bot }^{2})\varphi $, with two different ``diffusion'' coefficients for
the parallel and transverse subspaces, a serious question arises. Should we
expect both, or just one, of these to vanish as $T$ approaches criticality?
For the equilibrium system, we recall that the lowering of $\tau $ is a
consequence of the presence of interparticle interactions which slow (or
even reverse) the decay of density gradients, as $T$ decreases. In the
two-temperature model, the parallel subspace is held at a higher temperature
than the transverse one, suggesting that density gradients should decay much
faster in the parallel directions. Thus, we anticipate that, generically, $%
\tau _{\Vert }>$ $\tau _{\bot }$ so that criticality, in particular, is
marked by $\tau _{\bot }$ vanishing at positive $\tau _{\Vert }$. This is
borne out by the structure of typical ordered configurations, namely, single
strips aligned with the parallel direction, indicating that
``antidiffusion'', i.e., $\tau _{\bot }<0$, dominates in the transverse
directions below $T_{c}$.

The inequality satisfied by the diffusion coefficients is obviously crucial
for the correct description of criticality. Otherwise, it is fortunate that
the detailed dependence of the parameters in Eqns (\ref{TTa}) and (\ref{TTb}%
) on the microscopic $J$, $T$, and $T^{\prime}$, is not important. In the 
{\em disordered} phase, Eqns (\ref{TTa}) and (\ref{TTb}) can be simplified
even further, by truncating the Landau expansion after linear terms,
considering, in effect, a purely Gaussian theory.

To conclude this section, let us check whether Eqns (\ref{TTa},\ref{TTb})
satisfy the FDT. Comparing our noise correlations with Eqn (\ref{GN}), we
read off ${\Bbb N}=-\sigma _\alpha \delta _{\alpha \beta }{\partial }_\alpha
\partial _\beta $. It is straightforward to see that it is impossible to
recast the right hand side of Eqn (\ref{TTa}) in the Hamiltonian form of Eqn
(\ref{FDT}). We conclude that our continuum theory for the two-temperature
model generically {\em violates} the FDT. However, in Section 4 we will show
that the critical properties are controlled by a fixed point theory which is
associated with, surprisingly, a system (not Ising) 
in equilibrium \cite{S-epl}. The
implications are that FDT is restored in the critical region, for the {\em %
leading} singular parts of thermodynamic quantities.

\section{Simulation Results}

This section will be devoted to the main results from Monte Carlo
simulations on the two dimension model specified in Section 2.1. In the
first sub-section, we study the two-point correlations in the disordered
phase, showing convincing evidence for power law decays in configuration
space. In the second part, extensive investigations of the critical
properties will be presented. We will use subscripts $x$ and $y$ for the
``parallel'' ($\Vert $) and ``transverse'' ($\bot $) directions,
respectively. In the remainder of this article, $T$ will be quoted in units
of the Onsager value $T_{o}$.

\subsection{Two-point Correlations Far Above Criticality}

For an Ising model in equilibrium, regardless of the dynamical aspects of
the system, two-point correlations 
\[
G(\vec{x})\equiv \left\langle s_{\vec{i}}s_{\vec{i}+\vec{x}}\right\rangle 
\]
display power law decay only at the critical point. However, for a system
driven into a steady state under non-equilibrium conditions, the asymptotic
behavior of $G$ will depend on the dynamics. In particular, for our model at 
{\em any} $T$ {\em above} the critical point, $G(\vec{x})$ decays as $r^{-2}$%
, where $r\equiv \left| \vec{x}\right| $. Of course, the associated
amplitude must be direction dependent, so that the integral over space is
not divergent. In general, a simple rescaling of one co-ordinate (say, $x$)
will bring it to the same form as an electrostatic potential of a quadrupole
(in $d=2$) \cite{glms,LRC-TT}, so that there are long range
positive/negative correlations in the longitudinal/transverse direction,
i.e., 
\begin{equation}
G(x,0)\rightarrow A/x^{2}\quad \text{and}\quad G(0,y)\rightarrow -A^{\prime
}/y^{2}  \label{Gpower}
\end{equation}
and $A$ and $A^{\prime }$ are positive amplitudes depending on $J$ and $T$.
Had we used a finite $T^{\prime }$, both would depend on the difference $%
\left( T^{\prime }-T\right) $ as well, in a manner that vanishes as $%
T^{\prime }\rightarrow T$. The presence of these amplitudes can be traced to
FDT violation in these systems, as we will see in Section 4.1. Here, we
provide some details of how this power law behavior can be observed {\em %
directly} in $G$, as well as the simpler, ``indirect'' method using the
Fourier transform of $G.$ 

\begin{figure}[tbp]
\vspace{0.75cm}
\begin{center}
\begin{minipage}{0.35\textwidth}
    \epsfxsize = 1.\textwidth  
	\epsfbox{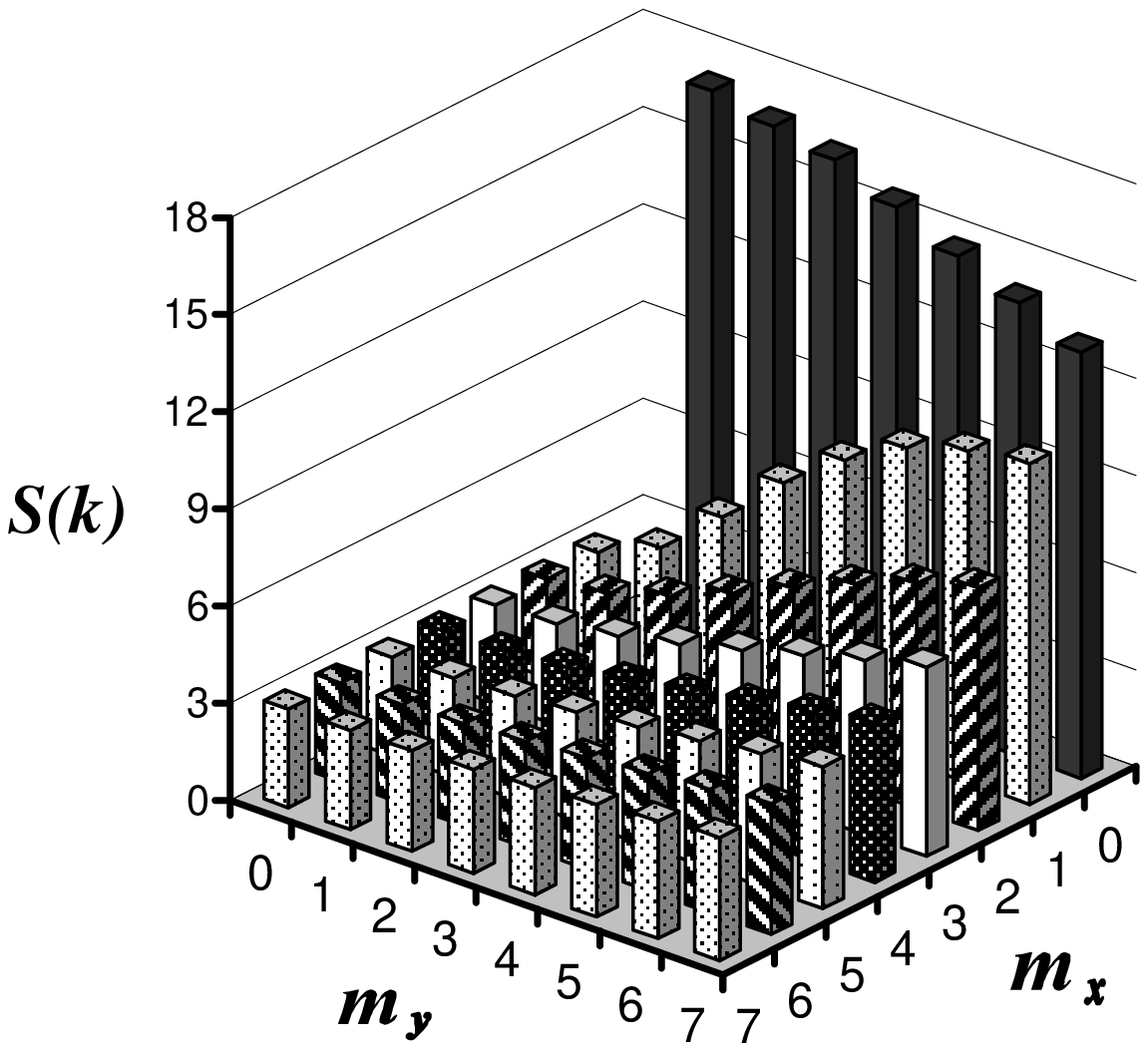} 
\end{minipage}
\end{center}
\caption{The structure factor $S(\vec{k})$ for a $128\times128$ lattice at
$T=2.0$.
The integers $\left(  m_{x},m_{y}\right)  $ are wavenumbers: 
$128\vec{k}/2\pi$.}
\end{figure}

We begin with the simpler way, i.e., by measuring the structure factor 
\[
S(\vec{k})\equiv \sum_{\vec{x}}e^{i\vec{k}\cdot \vec{x}}G(\vec{x})\,, 
\]
and seeking a discontinuity singularity \cite{zhls} at the origin $\vec{k}=0$%
. Unlike the Ornstein-Zernike form for the equilibrium Ising model, where 
\[
S_{eq}(\vec{k})\rightarrow \chi +O(k^{2}) 
\]
with $\chi $ being the static susceptibility, the limit depends on the
direction of $\vec{k}$ as it approaches the origin \cite{glms,zhls}. In
particular, the maximum range is embodied in, say, 
\begin{equation}
\lim_{k_{y}\rightarrow 0}S(k_{x}=0,k_{y})-\lim_{k_{x}\rightarrow
0}S(k_{x},k_{y}=0)\,.  \label{disc}
\end{equation}
Another convenient way to express such a discontinuity is 
\[
R\equiv \frac{\lim_{\left| k_{\bot }\right| \rightarrow 0}S(k_{\Vert }=0,%
\vec{k}_{\bot })}{\lim_{k_{\Vert }\rightarrow 0}S(k_{\Vert },\vec{k}_{\bot
}=0)}\neq 1\,. 
\]
Now, for a finite system, the wave-vectors $\vec{k}$ are discrete, being $%
(k_{x},k_{y})=2\pi (m_{x}/L_{x},m_{y}/L_{y})$ with integer $m_{i}$. Further,
our system is set at $M\equiv 0$, so that $S(0,0)\equiv 0$. Thus, through
measuring $S(\vec{k})$ at the two lowest wave-vectors: 
\[
\left( 2\pi /L_{x},0\right) \quad \text{and}\quad \left( 0,2\pi
/L_{y}\right) 
\]
in a series of systems with various $L_{x}$ and $L_{y}$, the discontinuity
singularity in (\ref{disc}) can be estimated by extrapolation. Here, we are
content to display the presence of such a discontinuity, as in the first
Monte Carlo study of driven systems \cite{kls}.

In Fig. 3.1, we show $S(\vec{k})$ for the $128\times 128$ case at $T=2.0$ .
For this plot, we made measurements every $20$ MCS on 5-10 runs, each up to $%
10^{6}$ MCS. Note the sizable difference between $S(2\pi /L_{x},0)$ and $%
S(0,2\pi /L_{y})$. By contrast, for the equilibrium case, these two
quantities would be the same, within statistical errors. 

\begin{figure}[tbp]
\vspace{0.5cm}
\begin{center}
\begin{minipage}{0.5\textwidth}
    \epsfxsize = \textwidth  
	\epsfbox{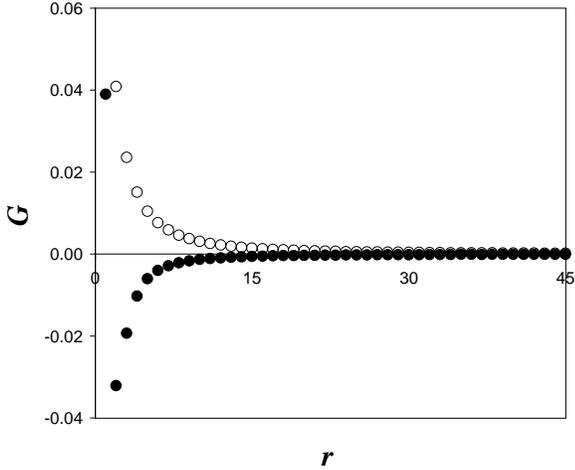}
    \vspace{-0.55cm}
\end{minipage}
\end{center}
\caption{The longitudinal (${\circ}$) and the 
transversal (${\bullet}$) correlation functions for a 
256 $\times$ 256 lattice, at $ T=3$.
To display data clearly, only $x,y \in [1,45]$ are shown.}
\end{figure}

\begin{figure}[tbp]
\begin{center}
\begin{minipage}{0.5\textwidth}
    \epsfxsize = \textwidth  
	\epsfbox{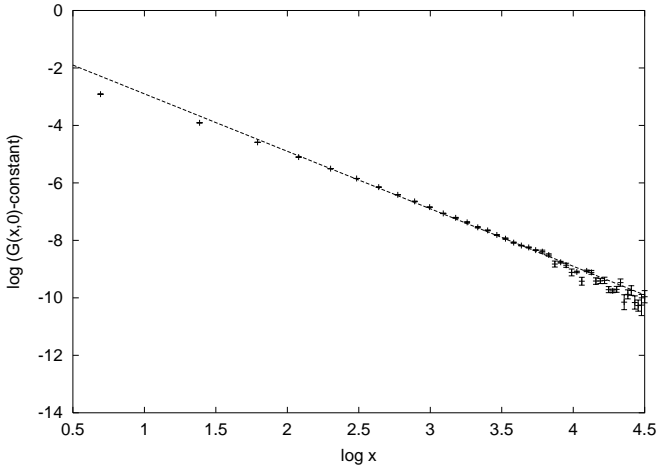}
\end{minipage}
\end{center}
\caption{Log-Log plot of $G(x,0)-constant$.The asymptote 
$x^{-2}$ is also shown.}
\vspace{-0.2cm}
\end{figure}

Parenthetically, let us point out that, for the $m_{x}=0$ series, a
multitude of long runs are crucial for minimizing the scatter in the data.
This ``difficulty'' may be traced to another manifestation of long range
correlations like Eqn (\ref{Gpower}), namely, the presence of strips along $%
x $ which are anti-correlated in $y$. This property favors slow decays of
strips with a range of widths, so that short runs may find only
configurations ``locked in'' particular regions of phase space. We believe
that this phenomenon is intimately tied to the metastability of ``multistrip
configurations'' first reported in \cite{VM}.

Though the discontinuity in $S(\vec{k})$ is the easiest way to deduce the
presence of a power law decay, it is naturally more satisfying to observe
the evidence directly. Indeed, in the early studies \cite{LRC-DDS,LRC-TT}, $%
G(\vec{x})$ is measured along the two axes and $\ln G$ vs. $\ln r$ is
plotted. However, the data used there to support the $r^{-2}$ behavior are
susceptible to criticisms. Typically, a straight line is drawn through less
than 10 scattered points, even though sizes up to $L_{x}=300$ were used. In
the study on the two-temperature model \cite{LRC-TT}, low statistics ($20$K
MCS) further limits the reliability. To remedy this situation, we carried
out a more careful study, measuring every $20$ MCS on 5-10 runs, each up to $%
10^{6}$ MCS, on a $256\times 256$ system at $T=3.0$. In Fig. 3.2, we present 
$G(x,0)$ and $G(0,y)$ vs. $x$ or $y.$ 

Apart from finding a more convincing evidence of the $r^{-2}$ decay, we
discovered a non-trivial finite size effect, which must be taken into
account before a log-log plot can be exploited. Due to the conservation law $%
M\equiv 0$, we have $\sum_{\vec{x}}G(\vec{x})\propto M\equiv 0$, so that the
asymptotic behavior of the two-point correlation in a {\em finite} system is
not zero. Indeed, for the totally disordered state ($T=\infty $), we must
have $G(\vec{x})=(N\delta _{\vec{x},\vec{0}}-1)/\left( N-1\right) $. As a
result, for high $T$, the power law tail (\ref{Gpower}) can be easily
overwhelmed by negative $O(1/N)$ contributions. In practice, the precise
values of these corrections (for finite $T$) are not known and represent the
only parameters in the fitting procedure. The result reveals excellent $%
r^{-2} $ decays. In Figs. 3.3 and 3.4, dashed lines of slope$\ -2$ are
included simply for the sake of comparison. From these plots, we conclude
that the behaviors (\ref{Gpower}) are well established.

\begin{figure}[tbp]
\vspace{0.5cm}
\begin{center}
\begin{minipage}{0.5\textwidth}
    \epsfxsize = \textwidth  
	\epsfbox{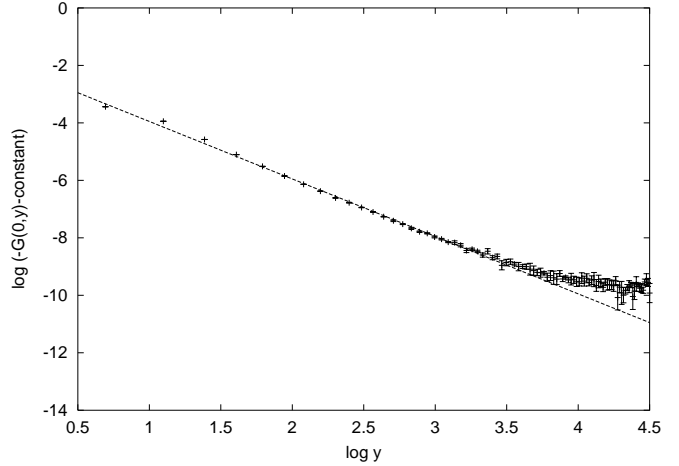}
\end{minipage}
\end{center}
\caption{Log-Log plot of $-G(0,y)-constant$.The asymptote 
$x^{-2}$ is also shown.}
\end{figure}

\subsection{Critical Properties}

As $T$ is lowered toward a critical value, $T_{c}$, correlations build up
and the system is expected to undergo a continuous transition to an ordered
state, characterized by phase segregation. Unlike in the equilibrium Ising
case, this segregation occurs {\em only} along the $y$-direction, a
phenomenon undoubtedly related to the anisotropic long-range correlation
above $T_{c}$. Another manifestation of this type of ordering is that {\em %
only} $S(0,2\pi /L_{y})$ diverges (as $O(N)$ for $T<T_{c}$), while $S(2\pi
/L_{x},0)$ remains $O(1)$ for all temperatures. The consequences of this
behavior are quite serious. Further, without some theoretical elucidations,
the pitfalls awaiting finite size analysis in computer simulation studies
are well hidden. As an example, in contrast to the equilibrium case, the use
of $L\times L$ samples (with a series of $L$'s) will not lead to consistent
collapse of data (Section 3.4). Instead, due to the conserved dynamics and
FDT violation, lengths/momenta in the two directions scale with different
exponents \cite{SZ-prl}, as will be shown in detail below (Section 4.1).
Thus, we resort to the method of ``strong anisotropic scaling''\cite
{ktl,wang}, i.e., using a series of {\em rectangular} samples with fixed $%
L_{x}/L_{y}^{1+\Delta }$ and non-trivial $\Delta .$

Before delving into the details of the results, let us provide an
``appetizer'' for the origins of a non-zero $\Delta $, from ``mean field
theory''. To describe this transition in the equilibrium case, Landau let $%
\tau $ (in Eqn \ref{HLG}) go through zero. As a result, the leading
relaxation term in (\ref{HLG}) vanishes, regardless of the noise ``matrix'' $%
{\Bbb N}$. In particular, for (\ref{MBa}) both diffusion coefficients vanish
at the same (critical) temperature. However, without the constraint of FDT,
it is not necessary that both vanish at one $T$. Indeed, to describe phase
segregation into {\em horizontal} strips (along with $S(2\pi /L_x,0)$ being $%
O(1)$), it is natural to let only the {\em transverse} diffusion
coefficient, $\tau _{\bot }$, in (\ref{TTa}), vanish. As a result, the
leading restoration terms become $-\nabla _{\bot }^4\varphi $ and $\partial
^2\varphi $, which translate into 
\[
-\partial _y^4\varphi \quad \text{and}\quad \partial _x^2\varphi \,, 
\]
in the language relevant for our simulations. Since the effects of these are
supposedly comparable in the critical region, the implication is 
\begin{equation}
k_x\sim k_y^2\,.  \label{kxky2}
\end{equation}
But, the lowest wavevectors available to our finite system are $(2\pi
/L_x,0) $ and $(0,2\pi /L_y)$. To insure that both types of fluctuations are
comparable in the critical region, we must choose $L_x\sim L_y^2.$ Thus, we
arrive at the ``mean field'' result 
\begin{equation}
\Delta _{MF}=1\,.  \label{MFDelta}
\end{equation}
As will be shown (Section 4.2), corrections to this result are expected to
be small, even in $d=2$ (unlike the uniformly driven lattice gas, where $%
\Delta =2$). Thus, for convenience, we simply used systems with $\Delta
\simeq 1$, namely, $20\times 20$, $45\times 30$, $80\times 40$, $125\times
50 $, and $180\times 60$. To have some confidence that $1-\Delta $ is indeed
small and positive, we also carried out runs with $120\times 50$ and $%
170\times 60$. All results are entirely consistent within statistical
errors, in that a good quality of data collapse is observed. By contrast,
the use of square samples {\em does not} lead to consistent data collapse (Section
3.2.3), signaling the presence of another scaling variable: 
\begin{equation}
a\equiv L_x/L_y^{1+\Delta }.  \label{s}
\end{equation}

Finally, let us remark on possible choices for order parameters. Due to the
conservation law, $M$ is fixed at zero and cannot serve as the order
parameter. Instead, in the seminal work on driven systems \cite{kls} and
several subsequent studies \cite{marro}, a quantity was used 
which is sensitive to
ordering into horizontal strips rather than vertical ones: 
\begin{equation}
\sum_{y}\left[ \sum_{x}s(x,y)\right] ^{2}-\sum_{x}\left[
\sum_{y}s(x,y)\right] ^{2}  \label{m*}
\end{equation}
However, (\ref{m*}) is likely to overestimate the segregation into
two macroscopic phases \cite{ktl,wang}. For example, it cannot distinguish
single strip configurations from multiple strip ones, as long as the latter
are all horizontal. Further, since it is a sum of $S(\vec{k})$ over all $%
\vec{k}$'s (along both axes in $k$ space) up to the ultraviolet cut-off, its
scaling properties are quite nebulous. Therefore, we decided upon the {\em %
structure factor} $S(0,2\pi /L_{y}),$ being the clearest measure of the
spontaneous symmetry breaking, as {\em our order parameter} \cite{ktl}.
Configurations with multiple strips contribute little to this quantity.
Also, as a correlation function associated with the smallest possible
wavevector, there is little doubt on how to compare simulation data with
field theoretic predictions.

\subsubsection{Crossing of a Cumulant Ratio, Data Collapse and the Exponent $%
\nu $}

For the equilibrium Ising model, Binder \cite{Binder} introduced a
convenient ratio $g(L,T)\equiv 3-\left\langle \left( \sum s\right)
^4\right\rangle /\left\langle \left( \sum s\right) ^2\right\rangle ^2$,
which varies monotonically between $2$ and $0,$ as $T$ ranges in $\left[
0,\infty \right] $, regardless of $L.$ In the thermodynamic limit, it should
be a simple step function, with the step occurring at the critical
temperature. For finite $L$'s, $g(L,T)$ curves should cross at temperatures
approaching $T_c.$ Furthermore, the value at the crossing is predicted to be
universal \cite{Binder}, so that this method has been the favorite for the
first determination of the critical temperature. 
For our
model, $\sum s$ is clearly futile, since it is identically zero. Instead, we
must consider the first non-zero wavenumber in the Fourier transform: 
\[
\tilde{s}\equiv \sum_{\vec{x}}e^{i\vec{k}\cdot \vec{x}}s_{\vec{x}}\quad 
\text{with}\quad \vec{k}=(0,2\pi /L_y)\,. 
\]
Note that we chose $\left\langle \left| \tilde{s}%
\right| ^2\right\rangle /\left( L_xL_y\right) $ as our order parameter. 
The equivalent cumulant
ratio here is \cite{ktl,wang} 
\begin{equation}
g(L_y,T)\equiv 2-\left\langle \left| \tilde{s}\right| ^4\right\rangle
/\left\langle \left| \tilde{s}\right| ^2\right\rangle ^2\,.  \label{g(T)}
\end{equation}
In this definition, the dependence on $L_x$ is suppressed, since we have
used only sizes which correspond to fixed $a\equiv $ $L_x/L_y^{1+\Delta }$.
Also, due to $\tilde{s}^{*}\left( \vec{k}\right) =\tilde{s}\left( -\vec{k}%
\right) \neq \tilde{s}\left( \vec{k}\right) $, it is more convenient to have 
$2$ as the first term, so that $g(L_y,\infty )=0$. At the opposite end, $%
g(L_y,0)=1$, so that $g(L_y\rightarrow \infty ,T)$ is precisely the
Heaviside function: $\Theta (T_c-T)$. In the inset of Fig. 3.5, we plot the
ratio against $T$ for the system sizes chosen. Within the errors of our
simulations, the crossing occurs at $T_c\simeq 1.37.$

\begin{figure}[htbp]
\begin{center}
\begin{minipage}{0.5\textwidth}
    \epsfxsize = 1.\textwidth  
	\epsfbox{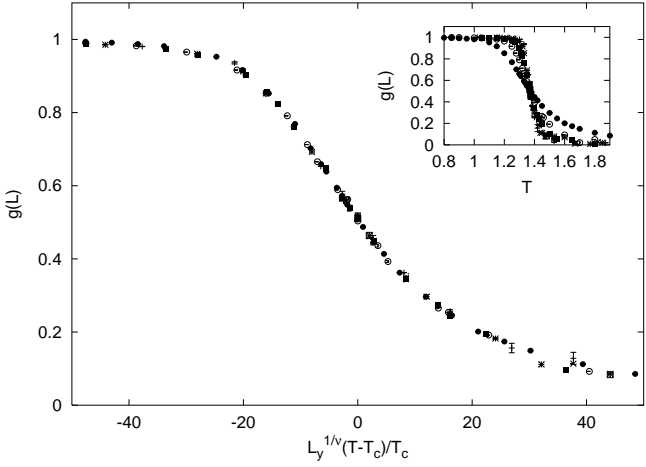} 
\end{minipage}
\end{center}
\caption{Scaling plot of $g(L_{y},T)$ vs.
$t{\equiv}L^{1/{\nu}}(T-T_{c})/T_{c}$,
for different system sizes $L_{x}{\times}L_{y}$: ${\mid}$ $ 180{\times}60$, ${\times} $ $ 170{\times}60$, $ {\star}$ $ 125{\times}50$, ${\square}$ $120{\times}50$, $ {\blacksquare}$ $ 80{\times}40$, ${\circ}$ $ 45{\times}30$, $ {\bullet}$ $ 20{\times}20.$
Inset:  $g$ vs. $T$.}
\end{figure}

More importantly, when plotted against the scaling variable 
\begin{equation}
\bar{t}\equiv \left( \frac{T}{T_{c}}-1\right) L_{y}^{1/nu}  \label{t}
\end{equation}
all curves may be collapsed into a single one (Fig. 3.5), with 
\begin{equation}
T_{c}\simeq 1.370(2)\,.  \label{Tc}
\end{equation}
and 
\begin{equation}
\nu \simeq 0.62(3)\,.  \label{nu}
\end{equation}
Note that we have tested the effects of having $\Delta $ slightly less than $%
1$ as well. We are encouraged by the data associated with these cases ($%
120\times 50$ and $170\times 60$) also lying within the typical scatter of
the ``universal'' curve. Thus, it is not crucial that we keep careful
account of the possible presence of a small $\left( 1-\Delta \right) $. The
value for the exponent$\ \nu $ is entirely consistent with the results of
renormalization group calculations (Section 4.2.3).

Finally, we emphasize that a general scaling ansatz for $g$ should be of the
form $g(L_{x},L_{y},T)=\bar{g}(a,\bar{t})$. Thus, the use of {\em square}
samples would have introduced the added complication of a varying $a$.
However, if $a$ were to enter $g$ as an additional power, then data collapse
can still be achieved, but with another value for $\nu $. We will return to
this topic in Section 3.2.3.

\subsubsection{Scaling Plot of $S(0,2\pi /L_{y})$ and the Exponent $\eta $}

From scaling arguments, the other independent exponent may be chosen as $%
\eta $. By definition, it is the anomalous dimension associated with the
structure factor at criticality in the limit of small $\vec{k}$. In models
with ``strong anisotropy'' such as ours, we must be more careful and define
it to be the anomalous dimension in $S(0,k_{y}\rightarrow
0;T=T_{c})\rightarrow k_{y}^{-2+\eta }$. (See Section 4.2.2 below for other $%
\eta $-like exponents.) For that reason, $S(0,2\pi /L_{y};T)$ serves
conveniently in a finite size scaling analysis to extract $\eta $. In
practice, it is more convenient to use \cite{ktl} 
\[
\tilde{S}\equiv S(0,2\pi /L_{y};T)\frac{L_{y}}{4L_{x}}\sin ^{2}\left( \frac{%
\pi }{L_{y}}\right) \,, 
\]
which is normalized to unity at $T=0$. Thus, the presence of scaling is
signaled by $\tilde{S}L_{x}L_{y}L_{y}^{-2+\eta } \sim \tilde{S}%
L_{y}^{\Delta +\eta }$ being a function of $\bar{t}$ alone. Of course, there
are generally two branches, associated with $T$'s above/below criticality.
For $T>T_{c}$, we may define $\gamma $ in the usual way, via $%
S(T)\rightarrow (T-T_{c})^{-\gamma }$ in the thermodynamic limit. In other
words, $\tilde{S}L_{y}^{\Delta +\eta }$should be proportional to $\bar{t}%
^{-\gamma }$. Using the definition of $\bar{t}$, we obtain the usual scaling
relation $2-\eta =\gamma /\nu $. Meanwhile, for $T<T_{c}$, $\tilde{S}$ plays
the important role of $\left\langle M^{2}\right\rangle $ for non-conserved
systems and serves to probe $\beta $, the order parameter exponent. So, $%
\tilde{S}L_{y}^{2\beta /\nu }$ should be a function of $\bar{t}$ alone,
leading to another scaling relation: $\Delta +\eta =2\beta /\nu $. Thus,
data collapse in a single scaling plot, for both $T$ above and below $T_{c}$%
, is a good test of the hyperscaling relation: 
\begin{equation}
2\beta +\gamma =\nu (2+\Delta )\,.
\end{equation}
As will be shown below, field theory predicts $\Delta =1-\eta /2$, so that
its validity can be tested by plotting $\tilde{S}L_{y}^{1+\eta /2}$ against $%
\bar{t}$ and checking for data collapse. Further consistency can be verified
by extracting the exponents $\gamma $ and $\beta $ from the two branches and
checking if they correspond to $(2-\eta )\nu $ and $(1+\eta /2)\nu /2$,
respectively. 

\begin{figure}[btp]
\begin{center}
\begin{minipage}{0.5\textwidth}
    \epsfxsize = 1.\textwidth  
	\epsfbox{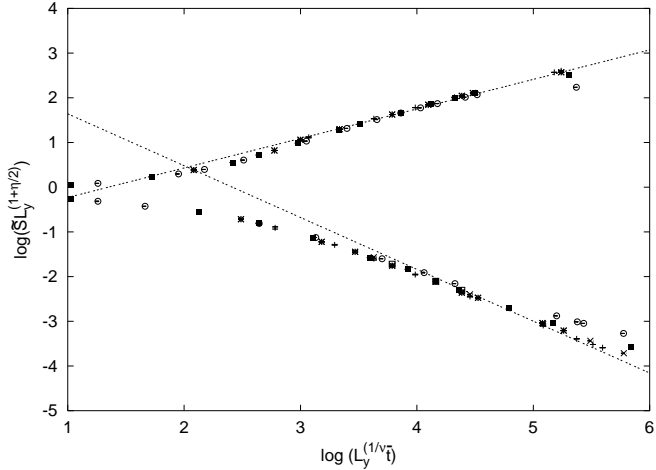} 
\end{minipage}
\end{center}
\caption{Scaling plot of
 $\tilde{S}L_{y}^{1+\eta /2}$ against $\bar{t}$ 
for different system sizes $L_{x}{\times}L_{y}$: 
${\mid}$ $ 180{\times}60$, ${\times} $ $ 170{\times}60$, $ {\star}$ $ 125{\times}50$, ${\square}$ $120{\times}50$, $ {\blacksquare}$ $ 80{\times}40$, ${\circ}$ $ 45{\times}30$.}
\end{figure}

Fig. 3.6 shows a log-log plot of the best data collapse of the two branches,
using (\ref{Tc}) for $T_{c}$ and (\ref{nu}) for $\nu $. To guide the eye,
and to show consistency, we have inserted lines with the expected slopes.
From this graph (and others with varying values of $\eta $) we estimate the
value 
\begin{equation}
\eta \simeq 0.13(4)\,\,.  \label{eta}
\end{equation}
and conclude that a consistent set of exponents exists, including 
\[
\beta \simeq 0.33(2)\,,\,\,\,\text{and}\,\,\ \gamma \simeq 1.16(6)\,. 
\]
The smaller samples deviate from the collapse as the temperature rises. This
deviation can probably be traced to the following. For small system sizes,
larger temperature ranges are needed to accumulate data for the scaling
plot, corresponding to $T$'s further outside the critical region.

\subsubsection{Data from Square Samples}

In order to test the importance of ``strong anisotropic scaling,'' we have
also performed simulations using {\em square} samples from $20\times 20$ to $%
80\times 80$. With data from such runs, it is possible to achieve reasonably
good collapse of $g(L\times L;T)$, using $T_{c}\simeq 1.34$ and $\nu \simeq
1.00$ (Fig. 3.7). However, with these values for $T_{c}$ and $\nu $, the best
``collapse'' we can achieve for $\tilde{S}$ is shown in Fig. 3.8, with $\eta
\simeq 0$. Better collapse of these data can be accomplished, but only by
using values of $T_{c}$ and $\nu $ {\em different} from those in Fig. 3.7,
or at the expense of a large, {\em negative }value of $\eta $! We believe
that the lack of good data collapse (with a consistent set of values for the
critical temperature and exponents) is a strong indication that another
scaling variable is playing a dominant role. In this case, it would be $%
L_{y}/L_{x}^{1+\Delta }$. If we use the mean-field value for $\Delta $, then
this variable ranges over a factor of $4$, for the set of data presented.
Such a large variation should be enough to explain the significant scatter
in Fig. 3.8.

\begin{figure}[htbp]
\begin{center}
\begin{minipage}{0.5\textwidth}
    \epsfxsize = 1.\textwidth  
	\epsfbox{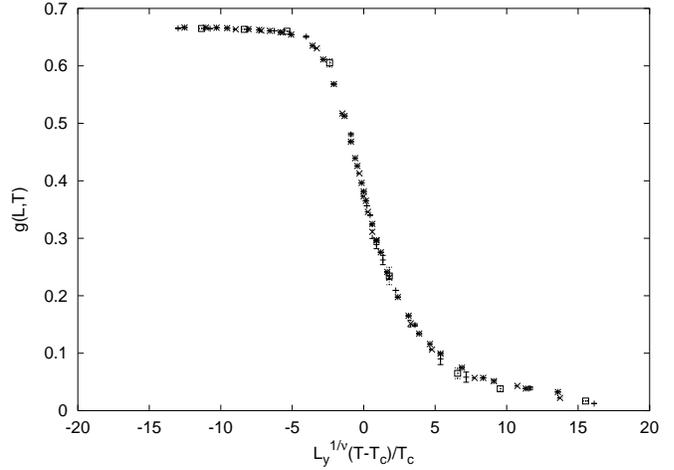} 
\end{minipage}
\end{center}
\caption{Scaling plot of $g(L{\times}L,T)$ vs. $(T/T_{c}-1)L^{1/{\nu}}$, 
with $T_{c}$=1.34 and $\nu$=1.00 
for different system sizes $L_{x}{\times}L_{y}$:
${\square}$ $80{\times}80$,
${\mid}$ $ 60{\times}60$, 
${\times} $ $ 40{\times}40$, 
$ {\star}$ $ 20{\times}20$.}
\end{figure}

\begin{figure}[htbp]
\begin{center}
\begin{minipage}{0.5\textwidth}
    \epsfxsize = 1.\textwidth  
	\epsfbox{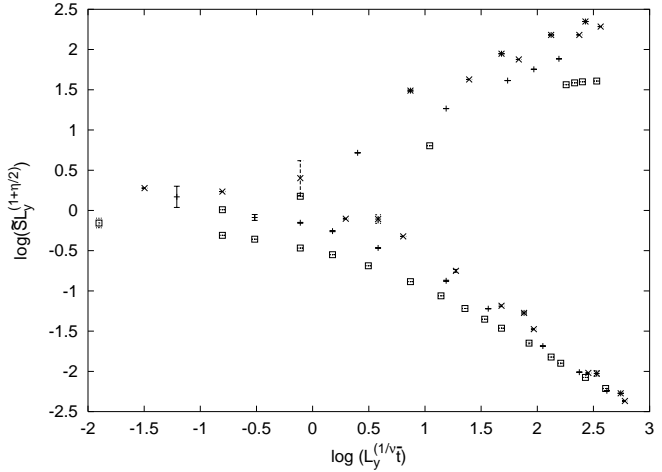} 
\end{minipage}
\end{center}
\caption{Scaling plot of
$\tilde{S}L_{y}^{1+\eta /2}$ against $\bar{t}$ 
for different system sizes $L_{x}{\times}L_{y}$: 
${\square}$ $80{\times}80$,
${\mid}$ $ 60{\times}60$, 
${\times} $ $ 40{\times}40$, 
$ {\star}$ $ 20{\times}20$.}
\end{figure}

In the literature, it has been claimed that consistent data collapse can be
achieved using square samples \cite{marro}. However, we believe that, by
using a different order parameter (Eqn (\ref{m*})), these authors may be
probing a different ``phase.'' In connection with the uniformly driven
lattice gas, the possible existence of such a ``new'' phase has been
conjectured recently \cite{stringy} and may resolve the differences observed
here. Further investigations are clearly necessary to clarify these issues.

\subsubsection{Fluctuations and Specific Heat}

For an Ising model in thermal equilibrium, both the free energy and the
internal energy, $\left\langle {\cal H}\right\rangle $, are well defined and
useful concepts. For non-equilibrium systems in steady state, the concept of
free energy becomes less clear, especially in our case of having thermal
reservoirs at two different temperatures. Nevertheless, we may continue to
regard $\left\langle {\cal H}\right\rangle $ as an internal energy,
associated with the interparticle interactions. Furthermore, since
anisotropy is expected, we may define the ``longitudinal'' and
``transverse'' parts of this energy, by $\left\langle {\cal H}%
_{x}\right\rangle $ and $\left\langle {\cal H}_{y}\right\rangle $
respectively, where 
\begin{equation}
{\cal H}_{x}\equiv -4J\sum_{\vec{i}}n\left( i_{x},i_{y}\right) 
n\left(i_{x}+1,i_{y}\right) \nonumber
\end{equation}
and
\begin{equation}
{\cal H}_{y}\equiv -4J\sum_{\vec{i}}n\left( i_{x},i_{y}\right) 
n\left( i_{x},i_{y}+1\right) \,.  \label{Hparts}
\end{equation}
Needless to say, these are nothing but the energies stored in the broken
bonds in the two directions.

As in equilibrium, the internal energy is a function of $T$ and we may
define its response to changes in $T$ as the heat capacity $C$

\[
C=\frac{\partial \left\langle {\cal H}\right\rangle }{\partial T}\,. 
\]
Similarly, its fluctuations can be measured. To emphasize the similarities
and differences between our system and one in equilibrium, we consider 
\[
C_{F}\equiv \frac{\left\langle \left( \Delta {\cal H}\right)
^{2}\right\rangle }{k_{B}T^{2}}\,. 
\]
Replacing ${\cal H}$ in these expressions by ${\cal H}_{x}$ and ${\cal H}%
_{y} $, we can define similar quantities: $C_{x},C_{Fx}$, etc. Of course,
had we used $T^{\prime }=T$, we would have $C=C_{F}=2C_{x}=2C_{Fx}\,$, etc.
Thus, measurements of all these quantities will provide not only a
specific-heat-like exponent $\alpha $, but also the degree of FDT-violation.

\begin{figure}[tbp]
\begin{center}
\begin{minipage}{0.5\textwidth}
    \epsfxsize = 1.\textwidth  
	\epsfbox{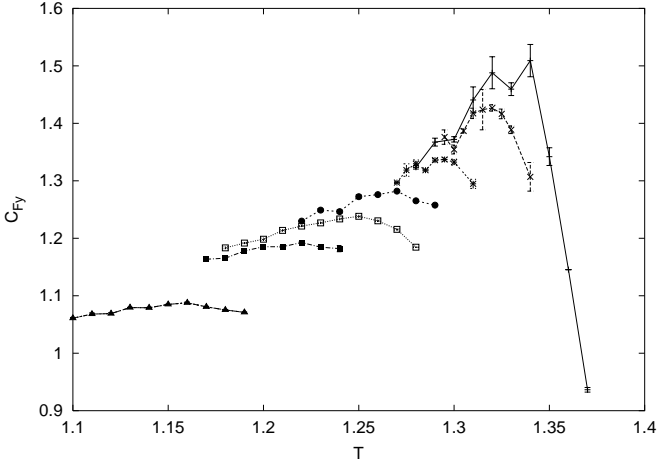} 
\end{minipage}
\end{center}
\caption{Fluctuations in ${\cal H}_y$, the ``transverse'' part of the 
energy, vs. $T$,
for different system sizes: 
${\mid}$ $ 180{\times}60$, 
${\times} $ $ 170{\times}60$, 
$ {\star}$ $ 125{\times}50$, 
${\bullet}$ $120{\times}50$, 
$ {\square}$ $ 80{\times}40$, 
${\blacksquare}$ $ 45{\times}30$, 
$ {\blacktriangle}$ $ 20{\times}20.$}

\end{figure}

We are able to compile reasonably good statistics on the fluctuations ($%
C_{F} $'s). For the critical region, the best data are associated with $%
C_{Fy}$, the fluctuations in the ``transverse'' part of the energy (Fig.
3.9). By contrast, due to the positive long range correlations, there are
fewer broken ``longitudinal'' bonds, which may be the reason behind the
lower quality of the data for $C_{Fx}$. To study scaling properties, we
consider first the {\em positions} of the peaks, $T_{p}(L_{y}).$ In Fig.
3.10 (inset), we show a log-log plot of $1-T_{p}/T_{c}$ vs. $L_{y}$, along
with a line of slope of $-1/0.62$, so that we have a consistent estimate for 
$\nu $. On the other hand, a similar plot for the {\em heights} of the peaks
(Fig. 3.10) leads us to $\alpha /\nu \simeq 0.28$. This value of $\alpha $
is also entirely consistent with the scaling relation predicted from field
theory (Section 4.2.3), i.e., $2-\alpha =\left( d+\Delta \right) \nu $.
Unfortunately, scaling plots for $C_{Fy}$ do not exhibit data collapse of
the same quality as those for $g$ and $S$. We believe that better statistics
would remedy this situation. Similarly, the data for energy are not precise
enough for estimates of the heat capacities ($C$'s). Nevertheless, we may
conclude from these observations that the {\em qualitative features} of the $%
C$'s are the same as the $C_{F}$'s. In particular, a log-log plot of the
heights of the $C_{y}$ maxima vs. $L_{y}$ provides us with a similar $\alpha 
$. However, it is not surprising that the numerical values of the $C$'s are
quite distinct from those for the fluctuations, since FDT is clearly
violated.

\begin{figure}[tbp]
\begin{center}
\begin{minipage}{0.5\textwidth}
    \epsfxsize = 1.\textwidth  
	\epsfbox{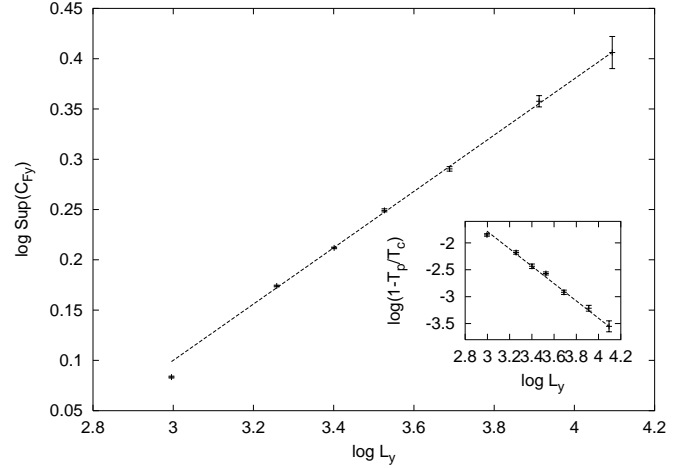} 
\end{minipage}
\end{center}
\caption{Log-log plots of the heights of the peaks (from Fig. 3.9) vs. $L_y$. 
Inset shows $ln(1-T_p/T_c)$ vs $ln L_y$, with $T_p$ being the positions 
of the peaks. The line is of slope $-1/0.62$.}
\end{figure}

\begin{figure}[bp]
\begin{center}
\begin{minipage}{0.5\textwidth}
    \epsfxsize = 1.\textwidth  
	\epsfbox{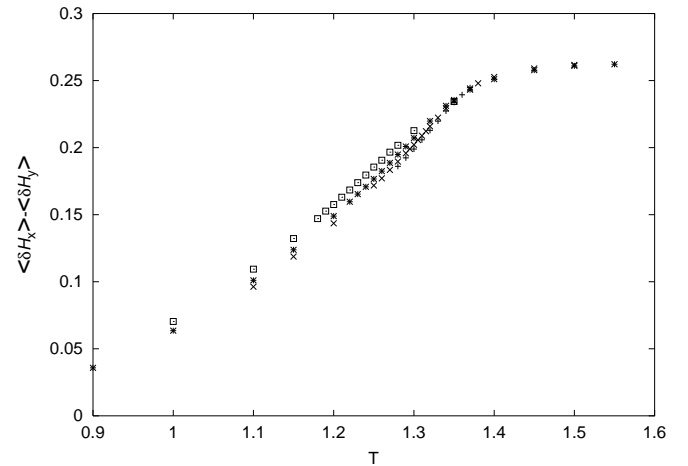} 
\end{minipage}
\end{center}
\caption{Energy flux 
$\left\langle \delta {\cal H}_x\right\rangle 
-\left\langle \delta {\cal H}_y\right\rangle $,
vs. T, for various system sizes:
$+$        $180 \times 60$, 
${\times}$ $125 \times 50$, 
${\star}$   $80 \times 40$, 
${\square}$ $45 \times 30$.}
\end{figure}

We end this section with a brief comment on energy {\em fluxes}. Since our
system is coupled to {\em two} temperature baths, we expect that there would
be a steady flow of energy through our system, from the bath with $T^{\prime
}$ to the one with $T.$ To be specific, with each spin-exchange update, we
can track whether it is associated with an $x$-bond or a $y$-bond.
Respectively, the energy change resulting from these exchanges would be
determined by
$T^{\prime }$ and $T.$ We have verified that, in the steady state, the 
{\em average} energy change associated with an $x$-bond spin-exchange is{\em %
\ positive. }Indeed, far from being just fluctuations, the average change in
a MCS, which we denote by $\left\langle \delta {\cal H}_x\right\rangle $, is
a few percent of $\left\langle {\cal H}\right\rangle $, for all the sizes we
investigated. Of course, being in steady state, the average change
associated with $y$-bond exchanges is of the same magnitude, but negative: $%
\left\langle \delta {\cal H}_y\right\rangle =-\left\langle \delta {\cal H}%
_x\right\rangle $.\ In other words, the energy flux through the system, $%
\left\langle \delta {\cal H}_x\right\rangle -\left\langle \delta {\cal H}%
_y\right\rangle $, is an extensive quantity. By contrast, for an equilibrium
system, the flux is fluctuating around zero, at the order of $\sqrt{N}$.
Similar differences have been observed in uniformly driven diffusive systems 
\cite{CCP99}. Qualitatively, this flux rises with temperature until about
the critical point. Thereafter, it appears relatively constant (Fig.~3.11). A
quantitative study of this flux, which lies outside the scope of this paper,
would undoubtedly yield rich information about an important non-equilibrium
characteristic of this system.

\section{Field Theoretic Studies}

Exact solutions for many-particle systems far away from thermal equilibrium
are restricted mostly to one-dimen\-sion\-al cases \cite{DL19} with excluded volume
interactions alone. Thus, progress for higher-dimensional models relies
entirely on mesoscopic continuum theories, such as the Lange\-vin equation for
our two-temperature model. While these field theories cannot be rigorously
derived from the microscopic dynamics, many of their key ingredients and
predictions are easily tested by Monte Carlo simulations. Considerable
confidence in these theories can therefore be established before they are
used to predict less easily measurable quantities. In the following, we will
outline how some of the simulations results of the preceding sections are
reflected in our continuum theory. We begin with the disordered phase,
before turning to fully developed critical behavior.

\subsection{Long-Range Correlations}

Well above $T_{c}$, typical configurations are nearly homogeneous, and
fluctuations of the local magnetization $\varphi(\vec{x},t)$ away from zero
are very small. It is therefore reasonable to neglect fourth-order gradient
terms as well as higher powers of $\varphi$ in our Langevin equation. The
resulting theory, for the equilibrium Ising model, is completely analytic
for all $T>T_{c}$. In contrast, for our far-from-equilibrium system, even
the disordered phase already exhibits nontrivial singularities. In the
following, we will discuss how these manifest themselves in our continuum
theory.

We begin with the simplified version of Eqns (\ref{TTa},\ref{TTb}) which is
fully appropriate for the disordered phase: 
\begin{equation}
\partial_{t}\varphi(\vec{x},t)=\lambda\{\tau_{\bot}\nabla_{\bot}^{2}\varphi+%
\tau_{\Vert}\partial^{2}\varphi\}-\vec{\nabla}\cdot\vec{\eta }{\bf (}\vec{x}%
,t)  \label{DPa}
\end{equation}
The noise correlations are given by 
\begin{equation}
\left\langle \eta_{\alpha}(\vec{x},t)\eta_{\beta}(\vec{x}^{\prime},t^{\prime
})\right\rangle =2\lambda\sigma_{\alpha}\delta_{\alpha\beta}\delta(\vec {x}-%
\vec{x}^{\prime})\delta(t-t^{\prime})\qquad.  \label{DPb}
\end{equation}
Since Eqn (\ref{DPa}) is linear, it is easily solved via a Fourier transform
to momentum and frequency space. With $\varphi(\vec{x},t) \linebreak 
= \int_{\vec{k},\omega}
\varphi(\vec{k},t)\exp i(\vec{k}\cdot\vec{x}+\omega t)$, where 
$\int_{\vec{k},\omega}\equiv\int\frac{d^{d}k}{(2\pi)^{d}}
\int\frac{d\omega }{2\pi}$, we find 
\begin{equation}
\varphi(\vec{k},\omega)=\frac{-i\vec{k}\cdot\vec{\eta}{\bf (}\vec{k},\omega)%
}{i\omega+\lambda\left( \tau_{\bot}k_{\bot}^{2}+\tau_{\Vert}k_{\Vert
}^{2}\right) +O(k_{\bot}^{4},k_{\Vert}^{4},k_{\bot}^{2}k_{\Vert}^{2})}
\label{phi}
\end{equation}
The corrections in the denominator are a reminder of the neglected
fourth-order derivatives. Arbitrary correlation functions follow from (\ref
{phi}) by averaging appropriate products of $\varphi(\vec{k},\omega)$ over
the Gaussian distribution of the noise.

The most remarkable property of the disordered phase is the presence of
power-law correlations, even well above $T_{c}$. In the equal-time structure
factor, being the Fourier transform of the correlation function, these
should be reflected as an anomaly near the origin. Starting from the full
dynamic structure factor, defined via $\left\langle \varphi (\vec{k},\omega
)\varphi (\vec{k}^{\prime },\omega ^{\prime })\right\rangle 
\linebreak
\equiv \delta
(\omega +\omega ^{\prime })\delta (\vec{k}+\vec{k}^{\prime })S_{0}(\vec{k}%
,\omega )$, its Fourier transform back into the time domain, $S_{0}(\vec{k}%
,t)$, is easily found: 
\begin{eqnarray}
S_{0}(\vec{k},t) &=& \left\langle \varphi (-\vec{k},t)\varphi (\vec{k}%
,0)\right\rangle _{0}  \nonumber \\
 &=& \frac{\sigma _{\bot }k_{\bot }^{2}+\sigma _{\Vert }k_{\Vert }^{2}}{\tau
_{\bot }k_{\bot }^{2}+\tau _{\Vert }k_{\Vert }^{2}+O(k_{\bot }^{4},k_{\Vert
}^{4},k_{\bot }^{2}k_{\Vert }^{2})} \nonumber \\
&\,& \times\exp \left[ -\lambda \left( \tau _{\bot
}k_{\bot }^{2}+\tau _{\Vert }k_{\Vert }^{2}\right) |t|\right]  \label{S(t)}
\end{eqnarray}
The subscript reminds us that we are considering a purely Gaussian theory.
Setting $t=0$ yields the {\em equal-time} structure factor, $S_{0}(\vec{k}%
)\equiv S_{0}(\vec{k},0)$: 
\begin{equation}
S_{0}(\vec{k})=\frac{\sigma _{\bot }k_{\bot }^{2}+\sigma _{\Vert }k_{\Vert
}^{2}}{\tau _{\bot }k_{\bot }^{2}+\tau _{\Vert }k_{\Vert }^{2}+O(k_{\bot
}^{4},k_{\Vert }^{4},k_{\bot }^{2}k_{\Vert }^{2})}\text{ .}  \label{S(0)}
\end{equation}
In contrast to what one might have expected naively, (\ref{S(0)}) is {\em not%
} a simple anisotropic generalization, such as $S_{e}(\vec{k})\propto
1/(\tau +k_{\bot }^{2}+bk_{\Vert }^{2})$, of the usual isotropic
Ornstein-Zernike structure factor $1/(\tau +k^{2})$ for the equilibrium
Ising model. The key difference resides in the discontinuity singularity
exhibited by (\ref{S(0)}).\ Defining 
\[
R\equiv \frac{\lim_{\left| k_{\bot }\right| \rightarrow 0}S(k_{\Vert }=0,%
\vec{k}_{\bot })}{\lim_{k_{\Vert }\rightarrow 0}S(k_{\Vert },\vec{k}_{\bot
}=0)}\text{ ,} 
\]
we find that both $S_{e}(\vec{k})$ and its isotropic version lead to $R=1$,
indicating that such an anisotropy is too weak to affect the long-wavelength
limit. In stark contrast, (\ref{S(0)}) results in $R=(\sigma _{\Vert }\tau
_{\bot })/(\sigma _{\bot }\tau _{\Vert })$, since the value of $S_{0}$ at
the origin depends on the angle under which $\vec{k}$ vanishes!

At a more fundamental level, this discontinuity singularity is a direct
consequence of the violation of the FDT. For linear theories such as ours,
the FDT demands that the matrices of noise correlations, $%
\sigma_{\alpha}\delta_{\alpha\beta}$, and of diffusion coefficients, $%
\tau_{\alpha}\delta_{\alpha\beta}$, should be proportional to one another.
Thus, if the FDT were to hold, the general form (\ref{S(0)}) would be
immediately reduced to $S_{e}(\vec{k})$. This is the case for, e.g., an
Ising model with anisotropic interactions. For our two-temperature model,
however, there is no such constraint, so that generically $R\neq1$. At
criticality, the FDT is maximally violated, i.e., $R$ {\em diverges} as $%
\tau_{\bot}\rightarrow0$.

One consequence of such a singularity in $S_{0}$ is that its Fourier
transform, the two-point correlation function $G(\vec{r})$, becomes {\em %
long-ranged}, decaying as $r^{-2}$ at large distances. The amplitude, in
addition to being proportional to $(R-1)$, has a dipolar angular dependence,
resulting in positive $G(x_{\Vert },\vec{x}_{\bot }=0)$ and negative $%
G(x_{\Vert }=0,\vec{x}_{\bot })$, in agreement with our data. Of course, our
continuum theory presupposes an infinite system, so that no finite size
corrections appear here. Moreover, the dipolar character combined with the $%
O(k^{4})$ corrections ensures that an appropriate angular average of $G$ is
again exponentially cut off \cite{DL17}. However, the divergence of $R$ at
criticality indicates a crossover to a different behavior, which will be the
subject of the next section.

To conclude, we note that the Ising ``up-down'' symmetry of our model
ensures that all {\em three-point} functions vanish identically above and at
criticality. This feature clearly distinguishes our model from the uniformly
driven system where such correlations are generically nonzero and can even
be singular at the origin, along selected directions \cite{3pf}.

\subsection{Critical Properties and the Nontrivial Fixed Point}

As the temperature approaches $T_{c}$, the local magnetization begins to
develop large fluctuations, both in absolute value and in gradients. This
signals the breakdown of the linear theory, Eqns (\ref{DPa},\ref{DPb}), as a
valid description of the disordered phase. We should therefore return to the
full Langevin equation, Eqns (\ref{TTa}) and (\ref{TTb}). Our first task is
to identify all couplings relevant for critical properties, i.e., retaining
non-zero values at the fixed point of the renormalization group. We will see
that simple dimensional analysis alone can already exhibit some of the key
differences between our two-temperature system and its equilibrium
counterpart, Model B. Most remarkably, the fixed point theory {\em restores}
the FDT, but with respect to a {\em different} Hamiltonian. We define a set
of critical exponents and perform a scaling analysis for our model, by
anticipating the scaling forms for a few key quantities. Finally, we provide
the technical foundation for this scaling analysis, in the form of an
explicit renormalization group calculation, up to and including two-loop
order \cite{S-epl}.

\subsubsection{The Fixed Point Theory and FDT Restoration.}

In the continuum theory, (\ref{TTa},\ref{TTb}), criticality is marked by the
vanishing of the transverse diffusion coefficient, $\tau_{\bot}$, while its
parallel counterpart $\tau_{\Vert}$ remains positive. Reconsidering our
Landau expansion in powers of gradients, we clearly need the $%
(\nabla_{\bot}^{2})^{2}\varphi$ term to limit fluctuations with large
transverse gradients, since $\tau_{\bot}\nabla_{\bot}^{2}\varphi$ cannot
serve this purpose near $T_{c}$. In contrast, fluctuations with large
parallel gradients are still controlled by $\tau_{\Vert}\partial^{2}\varphi$%
, so that the higher order term, $(\partial^{2})^{2}\varphi$, is not needed
and will be neglected. As a result, the two leading linear terms in the
Langevin equation, near criticality, are $(\nabla_{\bot}^{2})^{2}\varphi$
and $\tau_{\Vert}\partial^{2}\varphi$. Introducing a characteristic momentum
scale for {\em transverse} critical fluctuations, $|\vec{k}_{\bot}|\sim\mu$,
we conclude, on the basis of simple dimensional analysis, that {\em parallel}
wave vectors scale as $|k_{\Vert}|\sim|\vec{k}_{\bot}|^{2}\sim\mu^{2}$. In
the long-wavelength limit ($\mu\rightarrow0$) which is the focus of our
study, parallel gradients are therefore less relevant than transverse ones: $%
\partial^{2}\varphi^{3}$ may be neglected in favor of $\nabla_{\bot}^{2}%
\varphi^{3}$, and transverse noise correlations dominate parallel ones.
Similarly, the mixed term $\alpha_{\times}\partial^{2}\nabla_{\bot}^{2}%
\varphi$ is irrelevant compared to $(\nabla_{\bot}^{2})^{2}\varphi$.
Collecting so far, we arrive at a reduced Langevin equation, containing {\em %
relevant }terms only:

\begin{eqnarray}
\partial _{t}\varphi (\vec{x},t) &=& \lambda \{(\tau _{\bot }-\nabla _{\bot
}^{2})\nabla _{\bot }^{2}\varphi +\tau _{\Vert }\partial ^{2}\varphi 
+\frac{u}{3!}\nabla _{\bot }^{2}\varphi ^{3}\} \nonumber \\
&\,& +\zeta (\vec{x},t)  \label{FPLa}
\end{eqnarray}
Here, the noise has been restricted to the transverse subspace, with
correlations 
\begin{equation}
\left\langle \zeta (\vec{x},t)\zeta (\vec{x}^{\prime },t^{\prime
})\right\rangle =-2\lambda \sigma _{\bot }\nabla _{\bot }^{2}\delta (\vec{x}-%
\vec{x}^{\prime })\delta (t-t^{\prime })\qquad   \label{FPLb}
\end{equation}
This equation forms the starting point for the analysis of universal
critical properties. Continuing the dimensional analysis, we find that $%
\lambda t\sim $ $\mu ^{-4}$, characteristic for a conserved order parameter.
Care must be exercised with the scaling of the spatial volume. For example, $%
\delta (\vec{x}-\vec{x}^{\prime })\sim \mu ^{d+1}$, since parallel lengths
generate an extra factor $\mu ^{-1}$. By construction, both $\tau _{\Vert }$
and $\sigma _{\bot }$ are of order $1$. In fact, the latter is a trivial
coefficient which will be rescaled to $1$ in the following. The role of $%
\tau _{\Vert }$ is more intriguing and will be discussed in Section 4.2.3.
It is now easy to obtain $\zeta (\vec{x},t)\sim \mu ^{(d+7)/2}$ so that the
order parameter itself scales as $\varphi \sim \mu ^{(d-1)/2}$. The upper
critical dimension of our model is $d_{c}=3$ since $u\sim \mu ^{3-d}$. Since
all irrelevant couplings have been neglected, we refer to Eqns (\ref{FPLa},%
\ref{FPLb}), somewhat loosely, as the {\em fixed point theory}. Strictly
speaking, we should set $\tau _{\bot }=0$ also, since the naive scaling $%
\tau _{\bot }\sim \mu ^{2}$ indicates that zero is the fixed point value of
the transverse diffusion coefficient.

It is instructive to define $D=d+n$, where $n$ is the dimension of the
parallel subspace. In the spirit of the more general models alluded to in
Section 2.2, $n$ may be considered a model parameter. Of course, we require $%
d>n$, for a transverse subspace to exist. Then, the magnetization scales as $%
\mu^{(D-2)/2}$ and $u\sim\mu^{4-D}$, whence $D_{c}=4$. In this form, the
similarity to Model B scaling, where $n=0$, is quite apparent. However, we
emphasize that the {\em physical }upper critical dimension is $d_{c}=D_{c}-n 
$, {\em below} the Ising model value 4. This already indicates that the
two-temperature model falls {\em outside} the Ising universality class.
Clearly, the only ``interesting'' case is $n=1$: then, nontrivial exponents
are expected in {\em two} dimensions, as borne out by our Monte Carlo
simulations. In contrast, the choice $n=2$ results in an upper critical
dimension of $2$, for a model that is defined only in $d\geq3 $ dimensions.
As a result, only mean-field exponents should be observable in this case.

Returning to Eqns (\ref{FPLa},\ref{FPLb}), we observe an intriguing
consequence of neglecting irrelevant terms: the {\em reduced} Langevin
equation for the near-critical theory {\em obeys} the FDT. Defining ${\Bbb N}%
=-\lambda \nabla _{\bot }^2$, and noting that ${\Bbb N}$ is a positive
definite symmetric operator, we can rewrite our Langevin equation as follows:

\begin{equation}
\partial _t\varphi (\vec{x},t)=-{\Bbb N}\frac{\delta {\cal H}}{\delta
\varphi }+\zeta (\vec{x},t)
\end{equation}
with a Hamiltonian that is most easily expressed in momentum space: 
\begin{eqnarray}
{\cal H}\left[ \varphi \right] &=& \int_{\vec{k}}\frac 12\varphi (-\vec{k})%
\frac{k_{\bot }^2}{k_{\bot }^4+\tau _{\bot }k_{\bot }^2+\tau _{\Vert
}k_{\Vert }^2}\varphi (\vec{k})  \nonumber \\
&\,& + \frac u{4!}\int_{\vec{k}_1,\vec{k}_2,\vec{k}_3,\vec{k}_4}\varphi (\vec{k}%
_1)\varphi (\vec{k}_2)\varphi (\vec{k}_3)\varphi (\vec{k}_4) \nonumber \\
&\,& \times (2\pi )^d\delta (%
\vec{k}_1+...+\vec{k}_4)  \label{H}
\end{eqnarray}
Since the only relevant contribution to the noise acts in the transverse
subspace, its correlations are simply 
\begin{equation}
\left\langle \zeta (\vec{x},t)\zeta (\vec{x}^{\prime },t^{\prime
})\right\rangle =2{\Bbb N}\delta (\vec{x}-\vec{x}^{\prime })\delta
(t-t^{\prime })
\end{equation}

This structure leads to several important consequences. First of all, near
criticality, the static properties of the system are in fact
equilibrium-like, being controlled by the distribution $\exp (-{\cal H})$.
Interestingly, even though our {\em microscopic} dynamics is purely local,
effective {\em long-range} interactions are generated at the {\em mesoscopic}
level. These are dipolar in nature, as reflected by the static Gaussian
propagator 
\begin{equation}
S_{0}(\vec{k})=\frac{k_{\bot }^{2}}{k_{\bot }^{4}+\tau _{\bot }k_{\bot
}^{2}+\tau _{\Vert }k_{\Vert }^{2}}  \label{SP}
\end{equation}
In fact, it is helpful to recall the Hamiltonian for uniaxial ferromagnets
with dipolar interactions \cite{dipolar}. Assuming that the spins are
aligned with the parallel direction and located at $\vec{r}$ and $\vec{r}%
^{\prime }$, the dipolar interaction takes the form $s_{\vec{r}}U(\vec{r}-%
\vec{r}^{\prime })s_{\vec{r}^{\prime }}$, with $U(\vec{r})\propto \left[
d\,x_{\Vert }^{2}-r^{2}\right] /r^{d+2}$. A Fourier transform to momentum
space yields $U(\vec{k})\propto k_{\Vert }^{2}/k^{2}$. Including the
exchange interaction, we obtain the propagator for uniaxial dipolar
ferromagnets, $S_{d}(\vec{k})\propto \left[ \tau +k^{2}+bk_{\Vert
}^{2}/k^{2}\right] ^{-1}$, where $\tau \rightarrow 0$ marks the critical
point and $b$ is a constant. At first sight, the $\vec{k}$-dependences of
these propagators, $S_{d}(\vec{k})$ and $S_{0}(\vec{k})$, are obviously not
identical. However, closer examination reveals that the difference resides
in {\em irrelevant} terms alone (e.g., $k_{\Vert }^{2}$, in a sum with $%
k_{\bot }^{2}$)! Of course, the nonlinear term, in both cases, is just the
usual $\varphi ^{4}$ interaction. Thus, we conclude that the leading {\em %
static} critical singularities of our two-temperature model fall into the
universality class of an {\em equilibrium} Hamiltonian, which also describes
uniaxial dipolar ferromagnets. This result is quite remarkable, given that
we started from an FDT-violating microscopic dynamics. At a technical level,
it simplifies our renormalization group analysis, since the {\em statics} of
uniaxial dipolar ferromagnets has been discussed in the literature \cite{BZ}.

Second, the FDT generates a hierarchy of equations, relating correlation and
response functions. These are more easily discussed if we first recast our
theory in terms of a dynamic functional. The latter also represents a much
more convenient starting point for the RG calculation of the {\em dynamic}
critical properties. Introducing a Martin-Siggia-Rose \cite{MSR} response
field $\tilde{\varphi}(\vec{x},t)$, we obtain the dynamic functional at the
fixed point, 
\begin{equation}
{\cal J}[\tilde{\varphi},\varphi ]=\int dtd^dx\left\{ \tilde{\varphi}%
\partial _t\varphi +\tilde{\varphi}{\Bbb N}\frac{\delta {\cal H}}{\delta
\varphi }-\tilde{\varphi}{\Bbb N}\tilde{\varphi}\right\}  \label{DF}
\end{equation}
This form has the advantage that both correlation and response functions can
be computed as functional averages with weight $\exp (-{\cal J)}$.
Specifically, response functions are just expectation values containing the
field $\tilde{\varphi}(\vec{x},t)$. For example, $\left\langle \tilde{\varphi%
}(\vec{x}^{\prime },t^{\prime })\varphi (\vec{x},t)\right\rangle $
represents the response of the local magnetization $\varphi $, at $(\vec{x}%
,t)$, to an infinitesimal perturbation at $(\vec{x}^{\prime },t^{\prime })$.
If we were to add a magnetic field to the Hamiltonian ${\cal H}$, the
dynamic zero-field susceptibility would be just $\chi (\vec{k},\omega
)=k_{\bot }^2\left\langle \tilde{\varphi}(-\vec{k},-\omega )\varphi (\vec{k}%
,\omega )\right\rangle $ where the prefactor reminds us that the order
parameter is conserved and therefore cannot respond to uniform perturbations
(such as a homogeneous magnetic field). By virtue of the FDT, this
susceptibility is related to the dynamic structure factor, according to 
\begin{equation}
S(\vec{k},\omega )=2\lambda \omega ^{-1}\ Im \left[ \chi (\vec{k},\omega
)\right]  \label{FDT2}
\end{equation}
Higher order response and correlation functions satisfy similar relations.

Some caution must be exercised here. The FDT, and hence Eqn (\ref{FDT2}),
holds strictly only {\em at the fixed point} of the two-temperature model.
Away from the fixed point, the full functional is {\em generically
FDT-violating}, including irrelevant operators which cannot be recast in the
form prescribed by Eqn (\ref{DF}). Thus, these irrelevant operators generate
different corrections-to-scaling for response and correlation functions.
With respect to simulation data, this implies that (\ref{FDT2}) holds only 
{\em inside} the critical region, for the {\em leading singular parts} of $S$
and $\chi$. A similar statement qualifies the relation between energy
fluctuations and the specific heat. In this sense, ``FDT breaking''
operators are {\em dangerously} irrelevant: their absence mimics a symmetry
which is not strictly satisfied.

Before concluding this section, let us briefly return to the discussion of
more general models and consider the case of more than two temperatures, $%
T_{1}\geq T_{2}\geq...\geq T_{d}$. Naturally, we expect the {\em lowest} of
these to control criticality.
Let us assume that, say, the $m$ lowest temperatures are degenerate and are
decreased while keeping the remaining $n=d-m$ (higher) temperatures fixed.
In the continuum limit, this will give rise to a series of diffusion
coefficients, $\tau_{1}\geq\tau_{2}\geq...\tau_{n}>\tau_{n+1}=...=\tau_{d}$.
Not surprisingly, $\tau_{n+1},...,\tau_{d}$ will vanish first, defining the
transverse subspace. The remaining $n=d-m$ dimensions specify the parallel
subspace, marked by finite diffusion coefficients at criticality.

\subsubsection{Scaling Analysis}

Near criticality, large fluctuations on all length scales dominate the
behavior of the system, so that renormalization group techniques are
indispensable. These allow us to compute, e.g., scaling forms and critical
exponents explicitly, in an expansion about the upper critical dimension.
Here, we will anticipate the scaling properties of our model, deferring
technical details to the next section.

The discussion leading to the fixed point theory, Eqns (\ref{FPLa}) and (\ref
{FPLb}), suggests that the critical behavior of the two-temperature model is
distinct from its Ising origins: since parallel and transverse wave vectors
scale with different powers, the upper critical dimension is shifted to $%
d_{c}=3$. Anticipating renormalization, we reformulate this scaling as $%
|k_{\Vert }|\sim |\vec{k}_{\bot }|^{1+\Delta }$, introducing the {\em strong
anisotropy }exponent $\Delta $. For comparison, {\em weak} anisotropy refers
to systems where $\Delta $ remains zero and only scaling {\em amplitudes}
are anisotropic. Equilibrium examples for the former include, e.g., Lifshitz
points or structural phase transitions \cite{lifshitz}, while the Ising
model with anisotropic interactions falls into the second category. Far from
equilibrium, the usual driven lattice gas \cite{kls} is the prototype model
for strongly anisotropic scaling \cite{ktl,wang}. Since all characteristic
lengths are expected to scale with the same exponents near criticality, a
finite size scaling analysis should rely on aspect ratios with constant $%
L_{\Vert }/L_{\bot }^{1+\Delta }$.

For any system with strong anisotropy, irrespective of its universality
class, the renormalization group predicts the general scaling form of, e.g.,
the dynamic structure factor near criticality: 
\begin{equation}
S(\vec{k},t;\tau_{\bot})=l^{-2+\eta}S(k_{\Vert}/l^{1+\Delta},\vec{k}_{\perp
}/l,tl^{z};\tau_{\perp}/l^{1/\nu})  \label{SC}
\end{equation}

Here, $l$ is just a scaling factor. Eqn (\ref{SC}) can be viewed as a {\em %
definition} of the critical exponents $\nu$, $z$, $\eta$ and $\Delta$. The
latter is a new exponent, in addition to the usual two independent static
exponents $\nu$ and $\eta$, and the dynamic exponent $z$. Different
universality classes are distinguished by the characteristic values of these
exponents, expressed, e.g., through their $\epsilon$-expansions. These will
be discussed in the next section.

The scaling of the structure factor is particularly instructive, since
strong anisotropy plays a key role here. For example, four $\eta $-like
exponents can be defined \cite{DL17}, all of which would take identical
values in an isotropic system. In contrast, here we should introduce $\eta
_{\bot }$ and $\eta _{\Vert }$ via $S(k_{\Vert }=0,\vec{k}_{\perp },t=0;\tau
_{\perp }=0)\sim |\vec{k}_{\perp }|^{-2+\eta _{\bot }}$, and $S(k_{\Vert },%
\vec{k}_{\perp }=0,t=0;\tau _{\perp }=0)\sim |k_{\Vert }|^{-2+\eta _{\Vert
}}$, for $|\vec{k}|\rightarrow 0$. Similarly, the two-point correlation
function decays with exponents $\eta _{\bot }^{\prime }$ and $\eta _{\Vert
}^{\prime }$ for large distances, $G(r_{\Vert },\vec{r}_{\bot }=0,t=0;\tau
_{\perp }=0)\sim r_{\Vert }^{-(d-2+\eta _{\Vert }^{\prime })}$, and $%
G(r_{\Vert }=0,\vec{r}_{\bot },t=0;\tau _{\perp }=0)\sim r_{\Vert
}^{-(d-2+\eta _{\bot }^{\prime })}$. Fortunately, even though these four
exponents are generically different, they are not independent. Instead, they
are related to $\Delta $ and $\eta $ through simple scaling laws \cite{DL17}%
. For example, $\eta _{\bot }=\eta $, and $\eta _{\Vert }^{\prime }=\frac{%
\eta -\Delta (d-3)}{1+\Delta }$. Similarly, we can define two critical
exponents, $\nu _{\bot }$ and $\nu _{\Vert }$: choosing $l$ to satisfy $\tau
_{\perp }/l^{1/\nu }=1$, transverse momenta scale as $|\vec{k}_{\perp }|\sim
l\sim \tau _{\perp }^{\nu }\equiv \tau _{\perp }^{\nu _{\bot }}$, whence
parallel momenta obey $k_{\Vert }\sim l^{1+\Delta }\sim \tau _{\perp }^{\nu
\left( 1+\Delta \right) }\equiv \tau _{\perp }^{\nu _{\Vert }}$, resulting
in $\nu _{\bot }=\nu $ and $\nu _{\Vert }=\nu \left( 1+\Delta \right) $. The
scaling of the momenta with time $t$ determines two dynamic critical
exponents, $z_{\bot }$ and $z_{\Vert }$, according to $|\vec{k}_{\perp
}|\sim t^{-1/z_{\bot }}$ and $k_{\Vert }\sim t^{-1/z_{\Vert }}$. We easily
find $z_{\bot }=z$ and $z_{\Vert }=z/(1+\Delta )$. The dynamic exponents are
especially interesting here, since they have not been investigated
previously, unlike the static ones which agree with those for the dipolar
system \cite{BZ}.

To compare with simulation data, several other exponents are interesting.
The order parameter exponent $\beta$, obtained from an equation of state,
controls the response of the magnetization. In the absence of dangerous
irrelevant operators, it can be extracted from a scaling analysis, discussed
in the next section: 
\[
\beta=\frac{1}{2}\nu(d-1+\frac{\eta}{2}) 
\]

Other exponents of interest include the ``specific heat'' exponent $\alpha $%
. To be specific, let us consider fluctuations of the energy, $C_F$, first.
Taking the naive continuum limit of ${\cal H}$, the average total internal
energy would be identified as $\left\langle \int d^dx\,\varphi ^2(\vec{x}%
,t)\right\rangle $ (plus higher order, less singular terms). Thus, the
fluctuations are associated with the {\em connected} part of $\int
d^dxd^dy\left\langle \varphi ^2(\vec{x},t)\,\varphi ^2(\vec{y}%
,t)\right\rangle $. To extract the singular part of this operator, we only
need to consider the fixed point functional. But the latter corresponds to a
Hamiltonian system, so that the computation follows standard routes for
static systems. In this sense, there is no need to recompute $\alpha $,
since it will be related to $\nu $ through the usual scaling relation. The only
point we should emphasize is that, due to the anomalous scaling of the
longitudinal momenta ($|k_{\Vert }|\sim |\vec{k}_{\bot }|^{1+\Delta }\sim
\mu ^{1+\Delta }$), we have 
\begin{equation}
2-\alpha =\left( d+\Delta \right) \nu  \label{alpha}
\end{equation}
Since FDT is restored for the fixed point, we may also conclude that the
singular part of the heat capacity, $C$, is of the same form, so that there
will not be a new exponent. Of course, in contrast to an equilibrium system, 
$C$ is not strictly equal to $C_F$, so that different amplitudes associated
with the leading singularities may be expected.

We may further inquire into the singular parts of the energy ``difference'' $%
\left\langle {\cal H}_{x}\right\rangle -$ $\left\langle {\cal H}%
_{y}\right\rangle $ in simulations, which is identically zero in case the
two temperatures are equal. In higher dimensions, the equivalent quantity
would be $\left\langle {\cal H}_{\perp }\right\rangle -$ $\left( d-1\right)
\left\langle {\cal H}_{\parallel }\right\rangle $. Taking the naive
continuum limit, the corresponding operator is 
\begin{equation}
\left\langle \int \,\left\{\left( \nabla _{\bot }\varphi \right) ^{2}
-\left( d-1\right) \left( \partial\varphi \right) ^{2}\right\} \right\rangle 
\nonumber 
\end{equation} 
With a higher naive dimension
than the total energy, its singular parts should be of higher order, so that
verifying their presence would not be facile.. From the theoretical
perspective, perhaps a more interesting question is to explore the energy 
{\em flux:} $\left\langle \delta {\cal H}_{x}\right\rangle -\left\langle
\delta {\cal H}_{y}\right\rangle $, which is associated inherently with a
non-equilibrium process. However, since the fixed point dynamic functional
is Hamiltonian, we believe that the singular parts of such operators would
be irrelevant (though dangerously so). Though intriguing, the tasks of
deriving and analyzing their corresponding operators in field theory lies
outside the scope of the present paper.

\subsubsection{Renormalization Group Computation at Two-loop Order}

In this Section, we finally provide some technical details of the
renormalization group analysis. The starting point is the Langevin equation (%
\ref{FPLa},\ref{FPLb}), recast as (\ref{DF}), the dynamic functional \cite
{DF}. Following standard methods \cite{FT}, we perform a renormalized
perturbation expansion, organized in powers of the nonlinearity. We first
collect the elements of perturbation theory, i.e., the bare correlation and
response propagators and the vertex. We then focus on the one-particle
irreducible vertex functions and identify those which are primitively
divergent. Using dimensional regularization, we compute the associated poles
in $\epsilon \equiv d_{c}-d$, up to and including two loops, followed by the
renormalization of vertex functions and coupling constants. Finally, we
establish and solve the renormalization group equation for the vertex
functions, thus arriving at the full scaling behavior.

\begin{figure}[tbp]
\vspace{1.2cm}
\begin{center}
\begin{minipage}{0.5\textwidth}
    \epsfxsize = 1.\textwidth  
	\epsfbox{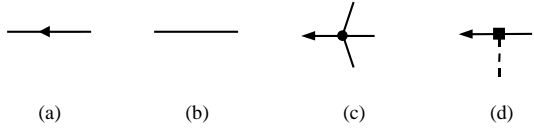} 
    \vspace{0.5cm}
\end{minipage}
\end{center}
\caption{Diagrammatic representation of the bare propagator (a),
correlator (b), vertex (c), and insertion (d).}
\end{figure}

The bare propagators are easily read off from the Gaussian theory,
corresponding to $u=0$, indicated by the subscript $\left\langle \cdot
\right\rangle _{0}$. For the response propagator, shown in Fig. 4.1a, we
find 
\begin{eqnarray}
G_{0}(\vec{k},\omega ) &\equiv& \left\langle \tilde{\varphi}(\vec{k},\omega
)\varphi (-\vec{k},-\omega )\right\rangle _{0} \nonumber \\
&=& \frac{1}{i\omega +\lambda
\left( \tau _{\bot }k_{\bot }^{2}+\tau _{\Vert }k_{\Vert }^{2}+k_{\bot
}^{4}\right) }  \label{RP}
\end{eqnarray}
while the correlation propagator (Fig. 4.1b), also referred to as the
correlator, is given by: 
\begin{eqnarray}
C_{0}(\vec{k},\omega ) &\equiv& \left\langle \varphi (\vec{k},\omega )\varphi (-%
\vec{k},-\omega )\right\rangle _{0} \nonumber \\
&=& \frac{2\lambda k_{\bot }^{2}}{\left|
i\omega +\lambda \left( \tau _{\bot }k_{\bot }^{2}
+\tau _{\Vert }k_{\Vert}^{2}+k_{\bot }^{4}\right) \right| ^{2}}  \label{CP}
\end{eqnarray}
The latter, of course, is just the dynamic structure factor for the Gaussian
theory. A comparison of Eqns (\ref{SC}) and (\ref{CP}) allows us to identify
the {\em Gaussian}, or {\em mean-field}, values of the critical exponents: 
\begin{equation}
\nu =\frac{1}{2},\quad z=4,\quad \eta =0\quad \text{and }\quad \Delta =1
\label{Gexp}
\end{equation}
Clearly, these will acquire nontrivial corrections of $O(\epsilon )$.

The theory possesses only a single nonlinearity, easily identified from Eqn (%
\ref{DF}) as $\lambda u\int dtd^{d}x(\nabla _{\bot }^{2}\tilde{\varphi}%
)\varphi ^{3}$. In Fourier space, this gives rise to a four-point vertex,
shown in Fig. 4.1c, 
\begin{equation}
-\lambda uk_{\bot }^{2}  \label{V}
\end{equation}
in Feynman integrals. Here, $k_{\bot }$ is the (transverse) momentum carried
by the $\tilde{\varphi}$-leg. However, the expansion is not organized simply
in powers of $u$, as one can infer from an additional scaling symmetry of
the theory: Due to the spatial anisotropy, parallel and transverse momenta
(or lengths) can be rescaled independently, namely $k_{\Vert }\rightarrow
k_{\Vert }/\alpha $ while $\vec{k}_{\bot }$ remains unchanged. The
functional, Eqn (\ref{DF}), is invariant under this transformation, provided
we rescale $\tau _{\Vert }\rightarrow \alpha ^{2}\tau _{\Vert }$ and $%
u\rightarrow \alpha u$, so that $u\tau _{\Vert }^{-1/2}$ is the invariant
form of the nonlinearity. For convenience, we also absorb a geometric factor 
$4\pi $ into the definition of the effective coupling constant: 
\begin{equation}
\tilde{u}\equiv \frac{1}{4\pi }u\tau _{\Vert }^{-1/2} \label{eff cc}
\end{equation}

In the following, we consider the {\em critical} theory, i.e., we set $\tau
_{\bot }$ to zero in all propagators. However, $\tau _{\bot }$ does require
renormalization which can be extracted from {\em insertions} of $\lambda
\tau _{\bot }\int dtd^{d}x(\nabla _{\bot }^{2}\tilde{\varphi})\varphi $ into
graphs of the critical theory. As usual, the insertion corresponds to a
two-point vertex, shown in Fig. 4.1d, carrying nonzero external momentum 
\cite{Pu}. For later reference, we note that vertex functions with
insertions can be resummed, to generate the vertex functions of the theory
above $T_{c}$ \cite{FT}.

Even though the detailed momentum dependence of propagators, vertex and
insertion is of course different from Model B, the topology and
combinatorics of the Feynman diagrams are
identical. This helps to simplify some of the technicalities. For example,
denoting bare one-particle irreducible vertex functions with $N$ ($\tilde{N}$%
) external $\varphi $- ($\tilde{\varphi}$-) legs and $L$ insertions as $%
\Gamma _{\tilde{N},N;L}$, it is straightforward to identify the primitively
divergent ones as $\Gamma _{1,1;0}$, $\Gamma _{1,3;0}$, and $\Gamma _{1,1;1}$%
. At the upper critical dimension, i.e., in $d=3$, the last two of these
both scale as $k_{\bot }^{2}$. This momentum dependence is already carried
by a factor of $k_{\bot }^{2}$ on one of the external legs, so that the
remaining integrals are momentum-independent. In contrast, $\Gamma _{1,1;0}$
has the dimension of $k_{\bot }^{4}$, so that the integral itself must
contribute another factor of $k_{\bot }^{2}$, in addition to the one carried
by the external leg. Similar to Model B, each of the three vertex functions
gives rise to one nontrivial renormalization, so that we anticipate {\em two}
independent exponents {\em provided} an infrared stable fixed point can be
found. All other vertex functions, including especially $\Gamma _{2,0;0}$,
are primitively convergent. It is now straightforward to evaluate these
diagrams and extract the singularities. Relegating the details to the
Appendix, we only quote the results: 
\begin{eqnarray}
\Gamma _{1,1;0}(\vec{k}_{\bot },0) &=&\,\lambda k_{\bot }^{4}\left\{ 1-\frac{%
1}{6}\tilde{u}^{2}k_{\perp }^{-2\epsilon }\hat{J}\right\} \label{BareGammas} \\
\Gamma _{1,3;0}(\{\vec{k}_{i},0\}) &=&-\lambda uk_{1,\perp }^{2}\left\{ 1-%
\frac{3}{2}\tilde{u}I_{1}+\frac{3}{4}\tilde{u}^{2}I_{1}^{2}
 +3\tilde{u}^{2}I_{2}\right\}   \nonumber \\
\Gamma _{1,1;1}(\vec{k}_{\bot },0;\vec{k}_{I},0) &=&-\lambda \tau _{\bot
}k_{\perp }^{2}\left\{ 1-\frac{1}{2}\tilde{u}I_{1}+\frac{1}{4}\tilde{u}%
^{2}I_{1}^{2}+\frac{1}{2}\tilde{u}^{2}I_{2}\right\}  \nonumber
\end{eqnarray}
Here, $\hat{J}$ as well as $I_{1}$ and $I_{2}$ are integrals defined in the
Appendix.

Next, we renormalize our theory, by introducing a set of renormalized
quantities (which are identified by the superscript $(R)$) and the
associated $Z$-factors. These are related to the original (bare) set of
quantities by: \ 
\begin{eqnarray}
\varphi &=& Z_{\varphi }^{1/2}\varphi ^{(R)}, \quad \quad \quad \tilde{\varphi}%
= Z_{\tilde{\varphi}}^{1/2}\tilde{\varphi}^{(R)}  \nonumber \\
\tau _{\Vert } &=& Z_{\varphi }^{-1}Z_{\tau _{\Vert }}\tau _{\Vert
}^{(R)}, \quad \  \tau _{\bot } = Z_{\varphi }^{-1}Z_{\tau _{\bot }}\tau _{\bot
}^{(R)}  \nonumber \\
\lambda  &=& (Z_{\varphi }/Z_{\tilde{\varphi}})^{1/2}Z_{\lambda }\lambda ^{(R)}
\label{Z-def} \\
u &=& \mu ^{\epsilon }Z_{\varphi }^{-2}Z_{u}u^{(R)}\quad \quad  \nonumber
\end{eqnarray}
Note that $u^{(R)}$ is dimensionless, and can be used to define a
renormalized effective coupling 
\begin{equation}
g\equiv \frac{u^{(R)}}{4\pi \sqrt{\tau _{\Vert }^{(R)}}}  \label{g}
\end{equation}
which is, unlike $\tilde{u}$, also {\em dimensionless}. Based on $\varphi
^{(R)}$ and $\tilde{\varphi}^{(R)}$, we consider a set of renormalized
vertex functions: 
\begin{eqnarray}
&\,& \Gamma _{\tilde{N},N;L}^{(R)}(\{\vec{k},\omega \};\lambda ^{(R)},\tau
_{\Vert }^{(R)},\tau _{\bot }^{(R)},g;\mu ) \nonumber \\
&\,& =Z_{\varphi }^{N/2}Z_{\tilde{%
\varphi}}^{\tilde{N}/2}\Gamma _{\tilde{N},N;L}(\{\vec{k},\omega \};\lambda
,\tau _{\Vert },\tau _{\bot },u)\,.  \label{NNLR}
\end{eqnarray}
The renormalization conditions consist of {\em demanding }that all $\Gamma
^{(R)}$'s be {\em finite} as $\epsilon \rightarrow 0$. As a result, through
Eqns (\ref{BareGammas}) and (\ref{Z-def}), all renormalized quantities are
implicitly dependent on $\mu $ -- the (momentum) scale at which these
conditions are imposed. Of the many routes to realize finiteness, the
minimal subtraction scheme is the most facile. Only pure poles in $\epsilon $
need to be introduced in the $Z$-factors, to cancel those in the $\Gamma $%
's. Thus, we can easily read off the results for the $Z$'s.

First, we find to{\em \ all orders in} $\epsilon $ that 
\begin{equation}
Z_{\varphi }Z_{\tilde{\varphi}}=Z_{\lambda }=Z_{\tau _{\Vert }}=1
\label{Z-comp-a}
\end{equation}
i.e., the couplings $\lambda $ and $\tau _{\Vert }$ require no
renormalization while the $Z$-factors for the fields are inverse to one
another. The remaining $Z$-factors must be determined order by order in
perturbation theory. To two loops, we obtain 
\begin{eqnarray}
Z_{\varphi } &=& 1+\frac{1}{6}\hat{J}g^{2}+O(g^{3})  \nonumber \\
Z_{\tau _{\bot }} &=& 1+\frac{1}{2}\hat{I}_{1}g+\left( \frac{3}{4}\hat{I}%
_{1}^{2}-\frac{3}{2}\hat{I}_{2}\right) g^{2}+O(g^{3})  \label{Z-comp-b} \\
Z_{u} &=& 1+\frac{3}{2}\hat{I}_{1}g+\left( \frac{15}{4}\hat{I}_{1}^{2}-3\hat{I}%
_{2}\right) g^{2}+O(g^{3})  \nonumber
\end{eqnarray}
where the expressions $\hat{I}_{1}$ and $\hat{I}_{2}$ are defined in the
Appendix.

Once the (finite) $\Gamma ^{(R)}$'s are known, we can find their scaling
behavior in the infrared limit by studying their ``flow'' as $\mu
\rightarrow 0$. Since the bare $\Gamma $'s are independent of $\mu $, flow
equations \cite{FT} for the $\Gamma ^{(R)}$'s arise from applying $\mu
\partial _{\mu }$, at {\em fixed} bare quantities, to Eqn (\ref{NNLR}): 
\begin{equation}
0=\left. \mu \partial _{\mu }\right| _{bare}\Gamma \propto \left\{ \left.
\mu \partial _{\mu }\right| _{g,...}+\beta (g)\partial _{g}+\ldots \right\}
\Gamma ^{(R)}  \label{CZE}
\end{equation}
where $\beta (g)\equiv \left. \mu \partial _{\mu }\right| _{bare}g$ and the
dots represent similar terms for the other parameters. Defering the complete
equation until below, we focus on $\beta (g)$, which controls the flow of
the effective coupling $g$. To describe critical behavior, we seek infrared
stable fixed points of the flow, i.e., zeroes of $\beta (g)$ which are
attractive under $\mu \rightarrow 0$. In our case, 
\begin{equation}
\beta (g)=-g\left\{ \epsilon -\frac{3}{4}\text{ }g+\frac{3}{4}Kg^{2}+\text{ }%
O(g^{3})\right\}  \label{beta}
\end{equation}
where $K\equiv 17/54+\ln \left( 2/\sqrt{3}\right) $. For $d>3$, the only
stable fixed point is the Gaussian $g^{*}=0$. It becomes unstable below the
critical dimension, i.e., in $d=3-\epsilon $, to be replaced by a second,
nontrivial fixed point which emerges smoothly from the origin: 
\begin{equation}
g^{*}=\frac{4}{3}\epsilon \ \left[ 1+\frac{4}{3}K\epsilon +O(\epsilon
^{2})\right] \text{ }  \label{fp}
\end{equation}
Near $g^{*}$, $g(\mu )$ flows as 
\begin{equation}
g(\mu )=g^{*}+\left[ g(\mu _{0})-g^{*}\right] \left( \mu /\mu _{0}\right)
^{\omega }  \label{beta-flow}
\end{equation}
if $\mu =\mu _{0}$ initially. Here, 
\begin{equation}
\omega =\epsilon -\frac{4}{3}K\epsilon ^{2}+O(\epsilon ^{3})  \label{omega}
\end{equation}
is {\em positive} (within perturbation theory), so that (\ref{fp}) is indeed
stable. This is also a critical exponent, controlling corrections-to-scaling.

Having established the existence of an infrared stable fixed point below the
upper critical dimension, we now focus on the scaling properties of our
theory which can be extracted from the RG equation. The latter expresses the 
$\mu $-independence of the bare vertex functions, at fixed bare parameters,
as a differential equation for the renormalized vertex function. To begin
with, let us introduce several additional Wilson functions: 
\begin{equation}
\gamma _{\bullet }\equiv \mu \partial _{\mu }Z_{\bullet }=\beta (g)\partial
_{g}\ln Z_{\bullet }\quad \text{and\quad }\kappa _{\bullet }\equiv \mu
\partial _{\mu }\ln (\bullet )\text{ }  \label{Wilson-def}
\end{equation}
where the symbol $\bullet $ represents any of the variables $\varphi $, $%
\tilde{\varphi}$, $\tau _{\Vert }$, $\tau _{\bot }$, or $\lambda $. Given
the definitions of the couplings and the $Z$-factors computed above, these
functions satisfy the following identities, valid to all orders in
perturbation theory: 
\begin{eqnarray}
\gamma _{\lambda } = \gamma _{\tau _{\Vert }}=0 \,, \quad \gamma
_{\varphi } &=& -\gamma _{\tilde{\varphi}} \,,
\quad \kappa _{\tau
_{\Vert }}=-\kappa _{\lambda }=\gamma _{\varphi } \,, \nonumber \\
\quad \kappa _{\tau _{\bot }} &=& \gamma _{\varphi }-\gamma _{\tau _{\bot }}\quad
\label{Wilson-a}
\end{eqnarray}
Thus, there are only two independent Wilson functions, $\gamma _{\varphi }$
and $\kappa _{\tau _{\bot }}$, whose expansions are easily computed up to $%
O(g^{2})$. For the following, only their values at the fixed point are
important: 
\begin{eqnarray}
\gamma ^{*} & \equiv & \gamma _{\varphi }(g^{*})\ =\frac{4}{3^{5}}\epsilon
^{2}+O(\epsilon ^{3})  \label{Wilson-b} \\
\kappa ^{*} & \equiv & \kappa _{\tau _{\bot }}(g^{*})=\frac{1}{3}\epsilon +%
\frac{2}{9}\left( \frac{3}{54}+K\right) \epsilon ^{2}+O(\epsilon ^{3})
\nonumber
\end{eqnarray}
Finally, we recall that the vertex functions $\Gamma _{\tilde{N},N;L}^{(R)}$
of the {\em critical} theory, with insertions of $\tau _{\bot }$, are easily
resummed to give the vertex functions $\Gamma _{\tilde{N},N\text{ }}^{(R)}$
of the disordered phase, at finite $\tau _{\bot }>0$. Suppressing the
superscript $(R)$ on all parameters for the sake of clarity, we quote the
full RG equation: 
\begin{eqnarray}
0 &=& \Big\{ \mu \partial _{\mu }+\beta (g)\partial _{g}+\tau _{\bot }\kappa
_{\tau _{\bot }}\partial _{\tau _{\bot }}+\tau _{\Vert }\kappa _{\tau
_{\Vert }}\partial _{\tau _{_{\Vert }}}+\lambda \kappa _{\lambda }\partial
_{\lambda }  \nonumber \\
&\,& -\frac{\tilde{N}}{2}\gamma _{\tilde{\varphi}}-\frac{N}{2}\gamma
_{\varphi } \Big\} \Gamma _{\tilde{N},N}(\{\vec{k},\omega \};\tau _{\Vert
},\tau _{\bot },\lambda ,g,\mu )  \label{RGE}
\end{eqnarray}
This equation can be solved, using the method of characteristics. In the
immediate vicinity of the fixed point, where the theory exhibits scaling, we
can incorporate dimensional analysis and the anisotropic scale
transformation into the solution, whence: 
\begin{eqnarray}
&\quad& \Gamma _{\tilde{N},N}(\{\vec{k},\omega \};\tau _{\bot })= \label{RG-sol-a} \\
&\,& =l^{p}\Gamma _{%
\tilde{N},N}(\{\vec{k}_{\bot }l^{-1},k_{\Vert }l^{-2+\gamma ^{*}/2},\omega
l^{-4+\gamma ^{*}}\};\tau _{\bot }l^{-2+\kappa ^{*}}) \nonumber
\end{eqnarray}
Here, the overall scaling exponent is 
\begin{equation}
p\equiv d+5-\frac{N}{2} ( d-1+\frac{1}{2}\gamma ^{*}) -\frac{%
\tilde{N}}{2} ( d+3-\frac{3}{2}\gamma ^{*} -\frac{3}{2}\gamma
^{*} )\quad .  \label{RG-sol-b}
\end{equation}
and the parameter $l$ is just a scaling factor, as in Eqn (\ref{SC}). In Eqn
(\ref{RG-sol-a}), we list only those parameters in the argument that take an
active part in the scaling form.

Let us illustrate how Eqn (\ref{SC}) emerges from the solution of the RG
equation. To obtain the scaling form of the structure factor, we recall the
connection between the two-point correlations and the two-point vertex
functions: 
\[
S(\vec{k},\omega ;\tau _{\bot })=\frac{\Gamma _{2,0}(\vec{k},\omega ,\tau
_{\bot })}{\left| \Gamma _{1,1}(\vec{k},\omega ;\tau _{\bot })\right| ^{2}} 
\]
Inserting the appropriate scaling forms, and performing a Fourier transform
from the time- into the frequency domain yields Eqn (\ref{SC}):
\begin{eqnarray} 
&\quad& S(\vec{k},t;\tau _{\bot })= \nonumber \\
&\,& = l^{-2+\gamma ^{*}}S(\vec{k}_{\bot
}l^{-1},k_{\Vert }l^{-2+\gamma ^{*}/2},tl^{4-\gamma ^{*}};\tau _{\bot
}l^{-2+\kappa ^{*}}) \nonumber 
\end{eqnarray}
In {\em analogy} with the scaling form of Model B, we now {\em define} two
static exponents 
\begin{equation}
\eta  \equiv \gamma ^{*}=\frac{4}{243}\epsilon ^{2}+O(\epsilon ^{3})
\end{equation}
and
\begin{eqnarray}
\nu &\equiv& \frac{1}{2-\kappa ^{*}} \nonumber \\
&=& \frac{1}{2}+\frac{1}{12}\epsilon +\frac{1}{36}\left[ \frac{67}{54}%
+\ln \frac{4}{3}\right] \epsilon ^{2}+O(\epsilon ^{3})
\end{eqnarray}
which are the only two independent indices. We emphasize that both of these
differ from their counterparts for the Ising model near $d=4$. Thus, the
two-temperature model falls into a different universality class, defined by
the dipolar system as shown by a comparison with the exponents quoted in 
\cite{BZ}. However, a two-loop calculation is necessary in order to exhibit
the differences between Model B and the dipolar system. At the order of one
loop, $\eta $ retains its mean-field value, and $\nu $ is determined by
combinatorics alone. Moving beyond statics, we find that the dynamic
exponent $z$ is related to the static ones by the conservation law: 
\[
z\equiv 4-\gamma ^{*}=4-\eta \quad . 
\]
While this scaling law is also obeyed by the Model B exponents, the explicit
result for $z$ is characteristic for the two-temperature model, as well as
the dipolar system in the presence of a conservation law for the
magnetization. Finally, we quote the strong anisotropy exponent $\Delta $,
which controls the anisotropic scaling of parallel and transverse momenta: 
\[
\Delta \equiv 1-\frac{1}{2}\gamma ^{*}=1-\frac{1}{2}\eta \quad . 
\]
Scaling laws relate all other exponents to these two, e.g., the ``specific
heat'' exponent $\alpha $ as discussed above.

The most interesting remaining index is the order parameter exponent $\beta $
which can be extracted from the equation of state. Since there are no
dangerous irrelevant operators here, a simple scaling analysis is
sufficient. In order to describe the effect of a magnetic field, we add a
term $\lambda \int dtd^{d}x\tilde{\varphi}\nabla _{\bot }^{2}h(\vec{x}%
_{\perp })$ to the dynamic functional. The transverse gradient operator
imposes the (relevant) effect of the conservation law. Clearly, this will
generate a nonvanishing, spatially varying magnetization, $M(\vec{k})$,
which breaks the Ising symmetry of our system. Introducing the vertex
generating functional $\Gamma \left\{ \tilde{\varphi},\varphi \right\} $,
the equation of state for the {\em renormalized} quantities follows from 
\[
h=\lambda ^{-1}\frac{\partial }{\partial k_{\bot }^{2}}\frac{\delta }{\delta 
\tilde{\varphi}}\Gamma \left\{ \tilde{\varphi},\varphi \right\} _{\left| 
\vec{k}=0,\tilde{\varphi}=0,\varphi =M\right. } 
\]
Here, the conservation law, reflected in the derivative $\partial /\partial
k_{\bot }^{2}$, selects the first nonzero Fourier component of the right
hand side. Expanding $\Gamma $ in powers of $M$, the right hand side is
reduced to a sum of vertex functions which are computed in the disordered
phase: 
\[
h=\lambda ^{-1}\frac{\partial }{\partial k_{\bot }^{2}}\sum_{N}\frac{M^{N}}{%
N!}\Gamma _{1,N}(\{\vec{k}_{\bot },k_{\Vert }=0,\omega =0\};\tau _{\bot
})_{\left| \vec{k}_{\bot }=0\right. } 
\]
The scaling form of the equation of state now follows directly from the
solution of the RG equation, Eqn (\ref{RG-sol-a}): 
\[
h(M,\tau _{\bot })=l^{(d+3-\frac{3}{2}\eta )/2} h(Ml^{-(d-1+\frac{1}{2}%
\eta )/2},\tau _{\bot }l^{-1/\nu }) 
\]
Choosing $l$ such that $Ml^{-(d-1+\frac{1}{2}\eta )/2}=1$, the equation of
state is recast in standard form 
\[
h(M,\tau _{\bot })=M^{\delta }f(\tau _{\bot }M^{-1/\beta }) 
\]
with an appropriately defined scaling function $f$. We can now read off the
magnetic-field exponent 
\[
\delta =\frac{d+3-\frac{3\eta }{2}}{d-1+\frac{\eta }{2}}\,, 
\]
and the order parameter exponent 
\[
\beta =\frac{1}{2}\nu (d-1+\frac{\eta }{2})\,. 
\]

\section{Summary and Outlook}

We have presented a detailed study of a non-equilibrium Ising system, with
Kawasaki exchange dynamics coupled anisotropically to two thermal baths.
Even when one of the baths is set at infinite temperature, this system is
known to display a second order phase transition similar to the equilibrium
model, with generic scale invariance at all temperatures above criticality 
\cite{glms,epl}. Both field theoretic renormalization group techniques and
Monte Carlo simulations on 2-dimensional square lattices have been used.
Excellent agreement between the two approaches is found for $T>T_{c}$. Since
this system displays strongly anisotropic scaling properties, we exploit
finite size scaling with {\em rectangular} samples to control the effects of
a new scaling variable associated with the aspect ratio (Eqn \ref{s}). The
results are entirely consistent with the predictions of the field theory. By
contrast, data from square samples fail to collapse, showing that the aspect
ratio indeed plays a significant role.

One of the more intriguing aspects of this system is the following. The
fixed point controlling critical properties can be associated with a
uniaxial magnet with dipolar interactions {\em in equilibrium}. More
specifically, the {\em leading} singularities in the thermodynamic functions
of our non-equilibrium lattice gas are, in the critical region, identical to
those for the equilibrium magnetic system. Of course, our lattice gas is
still a bona-fide non-equilibrium system. To find these aspects, we must
study either sub-leading singularities, such as corrections to scaling, or
quantities associated with dangerous irrelevant operators \cite{dangIRO}. A
good example of the latter is the energy flux through the system, which is
easily measured in simulations but quite difficult to compute theoretically.

Looking beyond this study, we see opportunities for further pursuit. Apart
from the obvious need for better statistics and larger system sizes, there
are many new avenues. To end this paper, we list a few. (a) Very little is
known about this system below the critical point. Only the fluctuations of
the interface were scrutinized \cite{KETTinterface}. Correlations within each
bulk phase are expected to be long-ranged, but confirmations from
simulations are still lacking. Similar to the case of a uniformly driven
lattice gas, domain growth during phase segregation is expected to display
strongly anisotropic time scales \cite{DDScoarsen}. Unlike that case, there
is no particle-hole asymmetry in our model. As a consequence, there may be
no serious disagreement between the microscopic lattice gas model and the
mesoscopic Cahn-Hilliard like approach. (b) For the driven lattice gas,
imposing open \cite{OBC} or skewed-periodic \cite{SPBC} boundary conditions
leads to surprisingly different behavior, especially at low temperatures.
Viable field theories for these systems are yet to be formulated. (c) On the
other hand, a complete renormalization group analysis exists for
multi-temperature models, as shown above. Though most of the predictions are
mean-field like, it would be interesting to carry out simulations to confirm
or disprove them.

Of course, this model represents only a small example in the vast field of
systems in non-equilibrium steady states. Other models subjected to two
thermal baths abound \cite{other TT}, not to mention complex physical
systems ranging from heating a pot on a stove to the ecosystem of the earth.
Hopefully, the richness displayed in this very simple two-temperature model
will spur further interest in pursuing other models and, most importantly,
enhance our insights into this vast class of physical systems in
non-equilibrium steady states.

\begin{center}
{\bf ACKNOWLEDGMENTS}
\end{center}
We thank A.D. Bruce, H.K. Janssen, O.G. Mouritsen, and Z. R\'{a}cz for many
illuminating discussions. The computational assistance of H. Larsen is
gratefully acknowledged. This research is supported in part by grants from
the US National Science Foundation under DMR-9727574, 
NATO under OUTR.CRG 961238, and the 
Danish Natural Science Research Council under 
Snr. 9901699. One of us (RKPZ)
thanks H.W. Diehl for his hospitality at the University of Essen, where some
of this work was performed.

\section{Appendix}

In this appendix, we include some details of how to extract the singular
parts of the three vertex functions of interest: $\Gamma _{1,1;0},\,\Gamma
_{1,3;0},$ and $\Gamma _{1,1;1}.$ Figs. 6.2-4 show the corresponding Feynman
graphs, up to and including two loops. Contributions that vanish in
dimensional regularization or due to causality have already been omitted. We
stress that we are only interested in the ultraviolet singular parts of the
vertex functions, in the critical theory. To regularize the infrared
behavior of our integrals, it is sufficient to keep{\em \ transverse}
external momenta finite. {\em Parallel} external momenta and external {\em %
frequencies} may eventually be set to zero, since they contribute to finite
terms only. The abbreviated notation $\int_{\vec{q},\omega }\equiv \int 
\frac{d^{d}q}{(2\pi )^{d}}\int $ $\frac{d\omega }{2\pi }$ will be used
throughout.

The structure of $\Gamma _{1,1;0}$ at this order is:

\begin{eqnarray}
&& \Gamma _{1,1;0}(\vec{k},\Omega ) = i\Omega +\lambda \left( k_{\bot }^4+\tau
_{\Vert }k_{\Vert }^2\right) 
 -\frac 12(\lambda u)^2k_{\bot }^2\times \nonumber \\
&\ & \times \int_{\vec{q}_1,...,\omega _3}q_{1\bot }^2G_0(%
\vec{q}_1,\omega _1)C_0(\vec{q}_2,\omega _2)C_0(\vec{q}_3,\omega _3) \nonumber \\
&\ & \times (2\pi)^{d+1}\delta (\sum_{i=1}^3\vec{q}_i-\vec{k})\delta (\sum_{i=1}^3\omega
_i-\Omega )  \label{110a}
\end{eqnarray}
Any neglected terms involve at least three loops. This expression can be
simplified considerably. First, it is straightforward to integrate over $%
\omega $'s, and the external $\Omega $ may be set to zero. Next, we absorb
the parallel diffusion coefficient $\tau _{\Vert }$ into the parallel
momenta, so that the effective coupling constant, Eqn (\ref{eff cc}),
appears explicitly. Finally, we can recast the remaining momentum integrals
in a more symmetric form, resulting in: 
\begin{equation}
\Gamma _{1,1;0}(\vec{k},0)=\lambda \left( k_{\bot }^4+\tau _{\Vert }k_{\Vert
}^2\right) -\frac 16\lambda \tilde{u}^2k_{\bot }^2J(\vec{k})\,,  \label{110b}
\end{equation}
where the remaining integral (which still carries a nontrivial $\vec{k}$%
-dependence, as we will see below) is defined as 
\begin{eqnarray}
&J&(\vec{k}) \equiv  \int_{\vec{q}_1,..,\vec{q}_3}\frac{q_{1\bot
}^2q_{2\bot }^2q_{3\bot }^2\;(2\pi )^d\delta (\sum_{i=1}^3\vec{q}_i-\vec{k})%
}{\left( q_{2\bot }^4+q_{2\Vert }^2\right) \left( q_{3\bot }^4+q_{3\Vert
}^2\right) \left( q_{1\bot }^4+q_{1\Vert }^2\right) }  \nonumber \\
&=& \int_{\vec{q}_1,\vec{q}_2,\vec{q}_3}S_0(\vec{q}_1)S_0(\vec{q}_2)S_0(\vec{q%
}_3) 
(2\pi )^d\delta (\sum_{i=1}^3\vec{q}_i-\vec{k})\,.  \label{Ja}
\end{eqnarray}
In the second line, we have recognized that the momenta under the integral
combine into three factors of the {\em static} Gaussian structure factor,
Eqn (\ref{SP}). Thus, the resulting expression, up to a prefactor $\lambda
k_{\bot }^2$, is {\em identical }to the {\em static} two-point vertex
function of the uniaxial dipolar ferromagnet, at two-loop order, with the
correct momentum integrals and combinatoric factors! This structure reflects
the restoration of the FDT, Eqn (\ref{FDT2}), and can be observed at each
order in perturbation theory.

Similar reductions occur for the other two vertex functions. Let us
demonstrate it for the one-loop graph of the (dynamic) four-point function, $%
\Gamma _{1,3;0}$. Its calculation is further simplified by choosing
transverse external momenta at a symmetry point, defined by 
\[
\vec{k}_{i\bot }\cdot \vec{k}_{j\bot }=k_{1,\perp }^{2}\left( 4\delta
_{ij}-1\right) /3\,, 
\]
for $i,j=1,...4$. Up to one loop, the expression reads: 
\begin{eqnarray}
&\Gamma& _{1,3;0}(\{\vec{k}_{i},\Omega _{i}=0\}) = -\lambda uk_{1,\perp
}^{2}+3(\lambda u)^{2}k_{1,\perp }^{2}\times \nonumber \\
&\ & \times \int_{\vec{q}_{1},...,\omega _{2}}q_{1\bot
}^{2}\;G_{0}(\vec{q}_{1},\omega _{1})S_{0}(\vec{q}_{2},\omega _{2}) \times
\nonumber \\
&\ & \times (2\pi
)^{d+1}\delta (\vec{q}_{1}+\vec{q}_{2}-\vec{k}_{1}-\vec{k}_{2})\delta
(\omega _{1}+\omega _{2})
\end{eqnarray}

\begin{figure}[tbp]
\vspace{0.4cm}
\begin{center}
\begin{minipage}{0.5\textwidth}
    \epsfxsize = 0.325\textwidth  
	\epsfbox{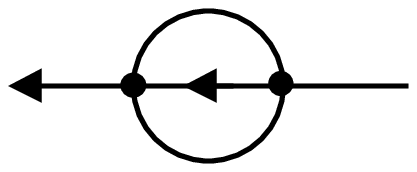} 
    \vspace{0.5cm}
\end{minipage}
\end{center}
\caption{Feynman diagram contributing to $\Gamma_{1,1;0}$.}
\end{figure}

Defining 
\begin{eqnarray}
&I&_{1}(\{\vec{k}_{i},0\}) \equiv  \int_{\vec{q}_{1},\vec{q}_{2}}\frac{%
q_{1\bot }^{2}q_{2\bot }^{2}\;(2\pi )^{d}\delta (\vec{q}_{1}+\vec{q}_{2}-%
\vec{k}_{1}-\vec{k}_{2})}{\left( q_{1\bot }^{4}+q_{1\Vert }^{2}\right)
\left( q_{2\bot }^{4}+q_{2\Vert }^{2}\right) }  \nonumber \\
&\ & =\int_{\vec{q}_{1},\vec{q}_{2}}S_{0}(\vec{q}_{1})S_{0}(\vec{q}_{2})(2\pi
)^{d}\delta (\vec{q}_{1}+\vec{q}_{2}-\vec{k}_{1}-\vec{k}_{2})  \label{I1}
\end{eqnarray}
we write 
\begin{equation}
\Gamma _{1,3;0}(\{\vec{k}_{i},0\})=-\lambda uk_{1,\perp }^{2}+\frac{3}{2}%
\lambda u\tilde{u}k_{1,\perp }^{2}I_{1}  \label{130a}
\end{equation}
Again, we recognize the zero-frequency part as the one-loop form of the {\em %
static} four-point function (up to a factor $\lambda k_{1,\perp }^{2}$),
involving just static propagators. The explicit reductions at the two-loop
level involve combinations of several dynamic diagrams into one static one.
For brevity, we only quote the results, so that the leading singular parts
of $\Gamma _{1,3;0}$ read: 
\begin{equation}
\Gamma _{1,3;0}(\{\vec{k}_{i},0\})=-\lambda uk_{1,\perp }^{2}\left\{ 1-\frac{%
3}{2}\tilde{u}I_{1}+\frac{3}{4}\tilde{u}^{2}I_{1}{}^{2}+3\tilde{u}%
^{2}I_{2}\right\}   \label{130b}
\end{equation}
where the last integral is given by 
\begin{eqnarray}
&I&_{2}(\{\vec{k}_{i},0\}) \equiv \int_{\vec{q}_{1},...,\vec{q}_{4}}
\prod_{i=1}^{4}S_{0}(\vec{q}_{i}) \times \label{I2} \\
&\ &\times(2\pi
)^{2d}\delta (\vec{q}_{1}+\vec{q}_{2}-\vec{k}_{1}-\vec{k}_{2})\delta (\vec{q}%
_{3}+\vec{q}_{4}-\vec{q}_{1}-\vec{k}_{3})
\nonumber 
\end{eqnarray}

Finally, we turn to the last vertex function, namely $\Gamma _{1,1;1}$ which
carries the insertion. As in Model B, the integrals contributing to $\Gamma
_{1,1;1}$ are just those of $\Gamma _{1,3;0}$: The insertion corresponds to
two of the four external legs being ``tied together''. In general, it
carries both momentum ($\vec{k}_I$) and frequency ($\omega _I$), but the
latter can be set to zero again. Thus, 
\begin{equation}
\Gamma _{1,1;1}(\vec{k},0;\vec{k}_I,0)=-\lambda \tau _{\bot }k_{\perp
}^2\left\{ 1-\frac 12\tilde{u}I_1+\frac 14\tilde{u}^2I_1^2+\frac 12\tilde{u}%
^2I_2\right\}  \label{111a}
\end{equation}

It is now straightforward, if tedious, to evaluate these integrals and to
extract the singularities. The one-loop diagrams present no difficulties. It
is rather instructive, however, to consider one of the nontrivial two-loop
diagrams as an example. We illustrate our procedure with the help of the
integral contributing to $\Gamma _{1,1;0}$: 
\begin{equation}
J(\vec{k})=\int_{\vec{q}_{1},\vec{q}_{2},\vec{q}_{3}}(2\pi )^{d}\delta
(\sum_{i=1}^{3}\vec{q}_{i}-\vec{k})\,S_{0}(\vec{q}_{1})S_{0}(\vec{q}%
_{2})S_{0}(\vec{q}_{3})\,.  \label{Jb}
\end{equation}
First, we integrate over the parallel momenta. To do so, we recast the $%
\delta $-function for the parallel momenta via its Fourier transform,
leading to a {\em product} of three
integrals of the form 
\begin{equation}
\int \frac{dq_{\Vert }}{2\pi }\frac{q_{\bot }^{2}}{\left( q_{\bot
}^{4}+q_{\Vert }^{2}\right) }\exp (ixq_{\Vert })=\frac{1}{2}\exp (-q_{\bot
}^{2}|x|)  \label{FT}
\end{equation}

\begin{figure}[tbp]
\vspace{0.7cm}
\begin{center}
\begin{minipage}{0.5\textwidth}
    \epsfxsize = 0.8\textwidth  
	\epsfbox{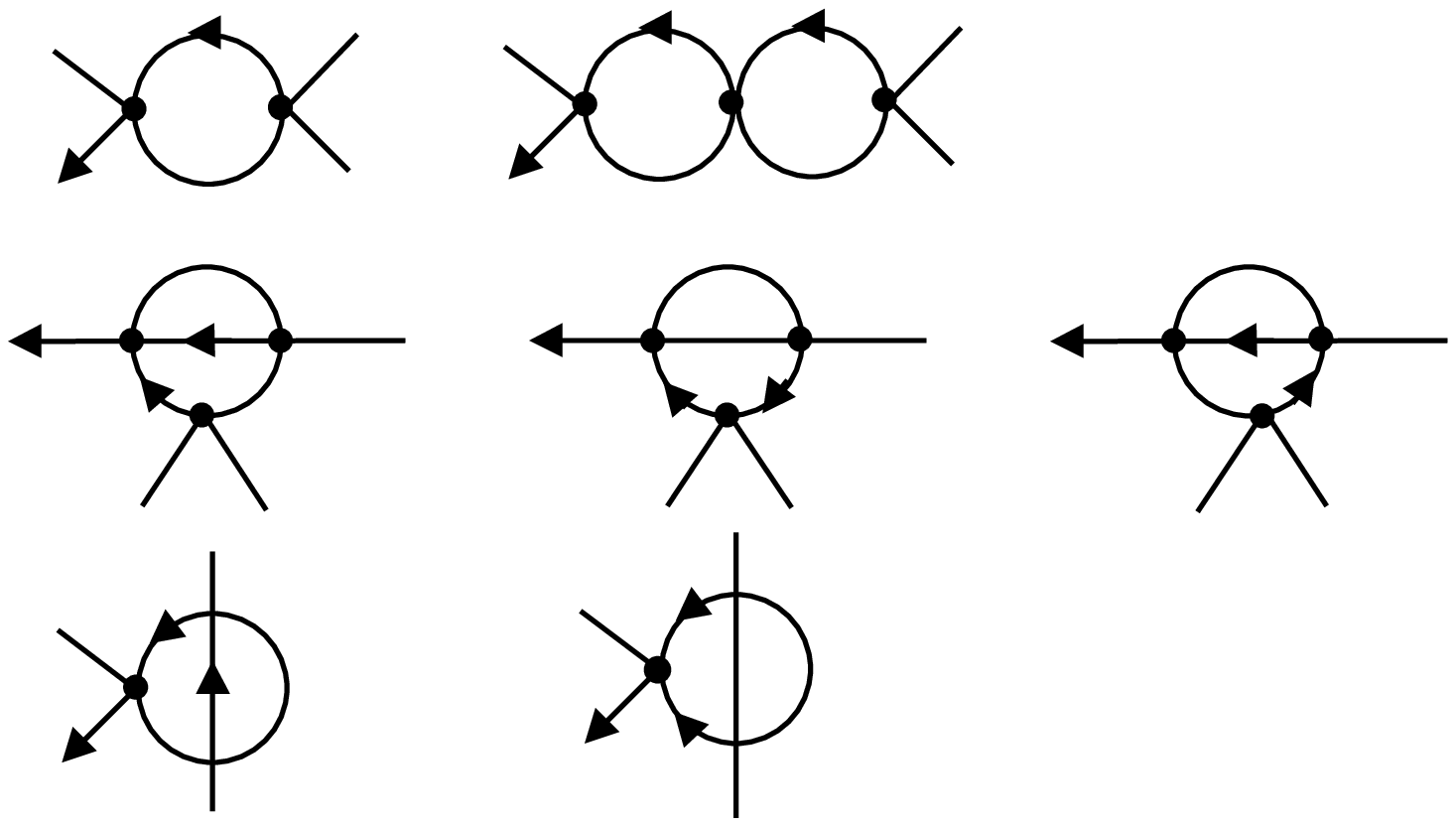} 
    \vspace{0.5cm}
\end{minipage}
\end{center}
\caption{Feynman diagrams contributing to $\Gamma_{1,3;0}$.}
\end{figure}

The integral over $x$ is now easily performed, and the external $k_{\Vert }$
may again be set to zero, resulting in 
\begin{eqnarray}
&J&(\vec{k}_{\bot },0)=\frac{1}{4}\int_{\vec{q}_{1\bot },\vec{q}_{2\bot },\vec{%
q}_{3\bot }}\frac{1}{\left( q_{1\bot }^{2}+q_{2\bot }^{2}+q_{3\bot
}^{2}\right) } \times \nonumber \\
&\ & \times (2\pi )^{d-1}\delta (\sum_{i=1}^{3}\vec{q}_{i\bot }-\vec{k}%
_{\bot })  \label{Jc}
\end{eqnarray}
The advantage is obvious: the extremely inconvenient $q_{\bot }^{4}$-terms
in the denominator have been reduced to quadratic ($q_{\bot }^{2}$) powers,
in the process cancelling the $q_{\bot }^{2}$-factors in the numerator. The
remaining integral is of a much simpler form and requires only the familiar
tools for evaluating standard $\phi ^{4}$-theory integrals \cite{FT}. In a
similar fashion, one can simplify the two-loop integrals for $\Gamma
_{1,3;0} $. Collecting our results for the integrals, we obtain, up to
$O(\epsilon )$ corrections: 
\begin{eqnarray}
J(\vec{k}_{\bot },0) &=&-\frac{1}{36\epsilon }(k_{\bot }^{2})^{1-\epsilon
}  \nonumber \\
I_{1}(\{\vec{k}_{i},0\}) &=&\frac{1}{2\epsilon }\left\{ 1+\frac{\epsilon }{2}%
\left[ -C_{E}+\ln (12\pi )\right] \right\} (k_{1,\bot }^{2})^{-\epsilon
/2} 
\nonumber \\
I_{2}(\{\vec{k}_{i},0\}) &=&\frac{1}{8\epsilon ^{2}}\left\{ 1+\epsilon
\left[ \frac{1}{3}-C_{E}+\ln (8\sqrt{3}\pi )\right] \right\} (k_{1,\bot
}^{2})^{-\epsilon }  \nonumber
\end{eqnarray}
Here, $C_{E}=0.577...$ is Euler's constant. The anticipated factor of $%
k_{\bot }^{2}$ is, finally, displayed explicitly in $J(\vec{k}_{\bot },0)$.
It indicates that this integral is a correction to the zero-loop term $%
\lambda k_{\bot }^{4}$ in Eqn (\ref{110b}): 
\begin{eqnarray}
\Gamma _{1,1;0}(\vec{k}_{\bot },0) &=&\lambda k_{\bot }^{4}-\frac{1}{6}%
\lambda \tilde{u}^{2}k_{\bot }^{2}\left\{ -\frac{1}{36\epsilon }(k_{\bot
}^{2})^{1-\epsilon }\left[ 1+O(\epsilon )\right] \right\} \nonumber \\
&=&\,\lambda k_{\bot }^{4}\left\{ 1-\frac{1}{6}\tilde{u}^{2}k_{\perp
}^{-2\epsilon }\hat{J}\right\} 
\end{eqnarray}

\begin{figure}[htbp]
\vspace{0.7cm}
\begin{center}
\begin{minipage}{0.5\textwidth}
    \epsfxsize = 0.8\textwidth  
	\epsfbox{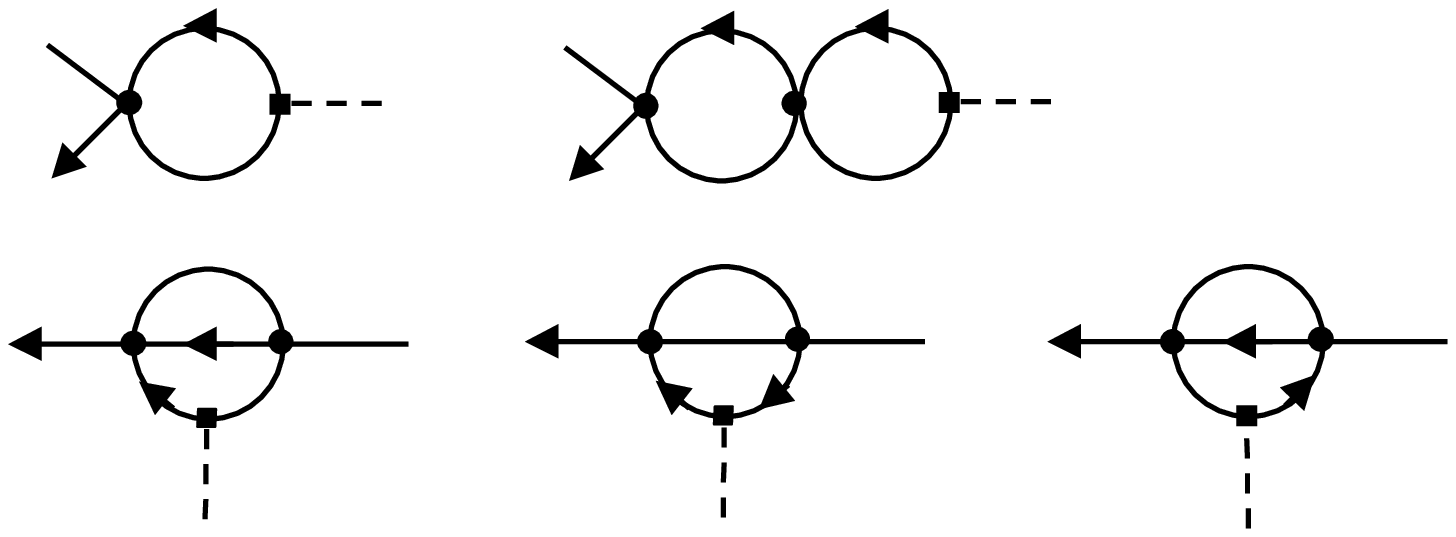} 
    \vspace{0.5cm}
\end{minipage}
\end{center}
\caption{Feynman diagrams contributing to $\Gamma_{1,1;1}$.}
\end{figure}

In the minimal subtraction scheme, only the pure $\epsilon $-poles in the $I$%
's and $J$ will be needed. Thus, we define 
\begin{eqnarray}
\hat{J} &=& - \frac{1}{36\epsilon }  \nonumber \\
\hat{I}_1 &=& \,\frac{1}{2\epsilon }  \nonumber \\
\hat{I}_1^2 &=& \,\frac{1}{4\epsilon ^2}\left\{ 1+\epsilon \left[ -C_E+\ln (12\pi
)\right] \right\}  \label{poles} \\
\hat{I}_2 &=& \,\frac{1}{8\epsilon ^2}\left\{ 1+\epsilon \left[ \frac 13-C_E+\ln
(8\sqrt{3}\pi )\right] \right\} \,,  \nonumber
\end{eqnarray}
where all $O(1)$ corrections have been neglected. Note that, within this
framework, $\hat{I}_1^2$ is ``more than the square of'' $\hat{I}_1$.


\begin{thebibliography}{99}
\bibitem{Pu}  Instead of introducing an additional ``field''
renormalization, $Z_{\tilde{\varphi}\varphi }$, for the insertion, we choose
a ``coupling constant'' renormalization of the parameter $\tau _{\bot }$.
Both procedures are equivalent.
\end{thebibliography}
\end{document}